\documentclass[twocolumn,preprintnumbers,amsmath,amssymb]{revtex4}

\usepackage{graphicx}
\usepackage{dcolumn}
\usepackage{bm}
\usepackage{graphics}
\usepackage[usenames]{color}
\usepackage[toc,page]{appendix}
\usepackage{hyperref}

\hypersetup{
    bookmarks=true,         
    unicode=false,          
    pdftoolbar=true,        
    pdfmenubar=true,        
    pdffitwindow=false,     
    pdfstartview={FitH},    
    pdftitle={My title},    
    pdfauthor={Author},     
    pdfsubject={Subject},   
    pdfcreator={Creator},   
    pdfproducer={Producer}, 
    pdfkeywords={keywords}, 
    pdfnewwindow=true,      
    colorlinks=true,       
    linkcolor=red,          
    citecolor=blue,        
    filecolor=magenta,      
    urlcolor=green           
}
\newcommand {\apgt} {\ {\raise-.5ex\hbox{$\buildrel>\over\sim$}}\ }
\newcommand {\aplt} {\ {\raise-.5ex\hbox{$\buildrel<\over\sim$}}\ }
\begin{document}

\preprint{}

\title{Pairing Fluctuations Determine Low Energy Electronic Spectra in Cuprate Superconductors}
\author{Sumilan Banerjee$^*$}
\author{ T. V. Ramakrishnan$^{*+}$}
\author{Chandan Dasgupta$^*$}
\affiliation{$^*$ Department of Physics, Indian Institute of Science, Bangalore 560012, India\\
$^+$ Department of Physics, Banaras Hindu University, Varanasi 221005, India}

\begin{abstract}
 We describe here a minimal theory of tight binding electrons moving on
the square planar Cu lattice of the hole-doped cuprates and mixed quantum mechanically with pairs of them (Cooper pairs).
Superconductivity occurring at the transition temperature $T_c$ is the long-range, $d$-wave symmetry phase coherence 
of these Cooper pairs. Fluctuations necessarily associated with incipient long-range superconducting
order have a generic large-distance behaviour near $T_c$. We calculate the spectral density of electrons
coupled to such Cooper pair fluctuations and show that features observed in Angle Resolved Photo Emission
Spectroscopy (ARPES) experiments on
different cuprates above $T_c$ as a function of doping and temperature emerge naturally in this description.  
These include `Fermi arcs' with temperature-dependent length and an antinodal pseudogap which fills
up linearly as the temperature increases towards the pseudogap temperature. Our results agree quantitatively
with experiment. Below $T_c$, the effects of
nonzero superfluid density and thermal fluctuations are calculated and compared successfully with some recent
ARPES experiments, especially the observed \emph{bending} or deviation of the superconducting gap from the
canonical $d$-wave form. 
\end{abstract}
\maketitle

\section{Introduction} \label{Sec.Introduction}
High-temperature superconductivity in hole-doped cuprates, accompanied by a `pseudogap phase' as well as
other strange phenomena, continues to be an outstanding problem in condensed matter physics for a quarter of
a century now \cite{PALee,JRSchreiffer,KHBennemann}. Over the years ARPES \cite{ADamascelli,JCCampuzano1} has uncovered 
a number of unusual spectral properties of electrons near the Fermi energy with definite in-plane momenta.
This low-energy electronic excitation spectrum is of paramount importance for explaining the rich and poorly 
understood collection of experimental findings \cite{PALee,JRSchreiffer,KHBennemann} from thermodynamic, transport and 
spectroscopic measurements on cuprate superconductors. We show here that the 
spectral function of electrons with momentum ranging over the \emph{putative} Fermi surface (recovered at high temperatures 
above the pseudogap temperature scale \cite{TTimusk,SHufner,MRNorman4}) is strongly affected 
by their coupling to Cooper pairs. On approaching $T_c$ i.e.~the temperature at which the Cooper pair phase
stiffness becomes nonzero, the inevitable coupling of electrons with long wavelength ($d$-wave symmetry) phase
fluctuations leads to the observed characteristic low-energy behavior observed in ARPES experiments. The idea that Cooper 
pair phase fluctuations are important in cuprate superconductivity has a long history; we mention only a few examples. The 
experimental realization that Cooper pair phase fluctuations are significant for a large regime of hole doping 
(below optimum doping) and temperature owes largely to the observation of 
large Nernst effect \cite{YWang} and enhanced fluctuation diamagnetism \cite{LLi} by Ong and coworkers. The same physics is 
implied in the early theoretical work \cite{VJEmery} of Emery and Kivelson.

\begin{figure}
\begin{center}
\begin{tabular}{c}
\includegraphics[width=9cm]{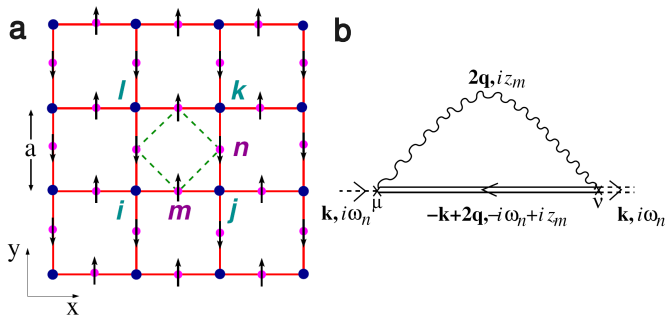}
\end{tabular}
\end{center}
\caption{{\bf Square lattice where cuprate superconductivity primarily resides and the pair fluctuation exchange process
for the self energy of electrons hopping on it.} {\bf a}, The blue circles at $\{\mathbf{R}_i\}$ represent the
 Cu atoms and the magenta ones at $\{\mathbf{R}_m\}$ (where $\mathbf{R}_m\equiv\mathbf{R}_{ij}\equiv\mathbf{R}_{i\mu}$ 
with $\hat{\mu}=\hat{x},\hat{y}$) represent bond centres (also the location of oxygen atoms). The arrows denote planar 
spins equivalent to $\psi_{ij}=\psi_{i\mu} \equiv \psi_m=\Delta_m \exp{(i\theta_m)}$ with $\Delta_m$ being the
length of the `spin' and $\theta_m$ the angle it makes with the $x$-axis; $d$-wave superconductivity translates
in spin language to $\mathrm{Ne'el}$ order as shown. {\bf b},  Feynman diagram\cite{GDMahan,AAltland} of the process in which an electron 
($\mathbf{k}$,$i\omega_n$) virtually becomes another electron
$(-\mathbf{k}+2\mathbf{q},-i\omega_n+iz_m)$ and absorbs a Cooper pair $(2\mathbf{q},iz_m)$ in the intermediate state.
The curly line denotes the Cooper pair propagator $D_{\mu\nu}(2\mathbf{q},iz_m)$ (here $z_m=2m\pi/\beta$ is
the bosonic Matsubara frequency, $m$ being an integer).}
\label{fig.BondLattice}
\end{figure}

The cuprates \cite{JRSchreiffer} (e.g. $\mathrm{La_{2-x}Sr_xCuO_4}$) exhibit superconductivity at unusually high 
temperatures on doping the parent compound, a Mott insulator, with holes ($x$ per
Cu site in the above case; Fig.\ref{fig.BondLattice}{\bf a} shows the square Cu lattice with lattice
spacing $a$). There is
long-standing evidence from the $(\cos k_xa-\cos k_ya)$ dependence of the superconducting gap $\Delta_\mathbf{k}$ on the 
in-plane momentum $\mathbf{k}$ of the
low energy electronic excitations (observed, for example, in ARPES experiments \cite{ADamascelli,JCCampuzano1}) 
that the pairing involves nearest neighbours on a tight binding
lattice. We therefore assume that the basic Cooper pair in hole-doped cuprates is the nearest-neighbour spin
singlet. Superconductivity is the long-range phase coherence of these Cooper pairs.    
The low energy degrees of freedom then are (the electrons and) the complex singlet pair amplitudes
\begin{eqnarray}
\psi_{ij}\equiv\frac{\langle b_{ij}\rangle}{\sqrt{2}}=\frac{1}{2}\langle a_{i\downarrow}a_{j\uparrow}-
a_{i\uparrow}a_{j\downarrow}\rangle\equiv \psi_m .
\end{eqnarray}
Here $a_{i\sigma}(a^\dagger_{i\sigma})$ destroys (creates) an electron with 
spin $\sigma$ at the lattice site $i$, sites $i$ and $j$
are nearest neighbours and the bond between them is uniquely labelled by the bond-centre site $m$ 
(see Fig.\ref{fig.BondLattice}{\bf a}). An interaction of the 
form $C\mathrm{Re}\left(\psi_m\psi^*_n\right)$ 
between nearest-neighbour bonds centered at $m$ and $n$, with positive $C$, leads to the observed $d$-wave 
symmetry superconductivity \cite{SBanerjee} below $T_c$. The crossover temperature $T^*(x)$ below 
which the equilibrium value $\langle|\psi_m|\rangle$ of the local pair amplitude becomes \emph{significant} is taken to be the 
`pseudogap' temperature. There is considerable experimental 
evidence for this view \cite{TTimusk,SHufner,MRNorman4}, though there is also 
the alternative view that $T^*(x)$ is associated with a new long-range order, e.g.~$d$-density wave \cite{SChakravarty}, time reversal symmetry breaking circulating currents
\cite{CMVarma}, electron nematic order\cite{SAKivelson}, stripes\cite{SAKivelson1} etc.

We work out here the effects of the quantum mechanical coupling between (fermionic) electrons and (bosonic)
Cooper pairs of the same electrons. On approaching $T_c$ from above, collective $d$-wave symmetry superconducting
correlations develop among the pairs with a characteristic superconducting coherence length scale $\xi$ which
diverges at the second-order transition temperature $T_c$. These correlations have a generic form at large distances
($>>$ the lattice spacing $a$). As we show here, the effect of these correlations on the electrons leads, for example, to 
temperature-dependent Fermi arcs \cite{MRNorman1} and to the filling of the antinodal pseudogap in the manner 
observed \cite{AKanigel1}. Further, the observed long-range order (LRO) below $T_c$ leads to a sharp antinodal spectral 
feature \cite{DLFeng,HDing} related to the nonzero superfluid density, and thermal pair fluctuations cause a deviation 
(`bending') of the inferred `gap' as a function of $\mathbf{k}$
from the expected $d$-wave form $(\cos{k_xa}-\cos{k_ya})$. The bending, being of thermal origin, decreases with decreasing 
temperature, in agreement with some ARPES measurements \cite{KTanaka,WSLee}.

 We describe the model used for our calculation in Section \ref{Sec.Model}, followed by a section (Section
\ref{Sec.Methods}) on the method of calculation and phenomenological inputs that are extracted from various
experiments. We have compared our results with a large number ARPES experiments, as mentioned in the preceding
paragraph. These are reported in Section \ref{Sec.Results}. There have been many previous 
studies of the effects of phase fluctuations on electrons; we mention a few of them in the discussion section
(Section \ref{Sec.Discussion}) along with some concluding remarks. Our attempt here is based on a particular
view of the superconducting transition, does \emph{not} assume preformed $d$-wave pairs, and a detailed
comparison with a large number of ARPES experiments is made. The present paper, in our view, represents 
a significant development, an extension of the approach of Ref.\cite{SBanerjee} to explicitly include electronic 
(fermion) degrees of freedom in addition to the Cooper pair (bosonic) ones. It
also shows that a simple minimal theory can account for for a variety of ARPES data, which have hitherto been analyzed
separately, using phenomenological \cite{MRNorman2} or problem specific \cite{MFranz,EBerg} models. 
Appendices describe some technical details of our calculations.

\section{Model} \label{Sec.Model}
A complete model for electrons in the square planar Cu lattice with relatively weak interlayer coupling and
with interactions which strongly affect their motion is needed for describing low energy cuprate phenomena
comprehensively. There is no such unanimously accepted description yet. We use a partial, phenomenological
model as follows. The electrons are described by the Hamiltonian,
\begin{eqnarray}
\hat{\mathcal{H}}&=&-\sum_{i,j,\sigma} \tilde{t}_{ij} a^\dagger_{i\sigma} a_{j\sigma} - \sum_{<ij>}
\tilde{J}_{ij} b^\dagger_{ij}b_{ij} \label{eq.Hamiltonian},
\end{eqnarray}
which consists of a quantum mechanical intersite hopping term, where $\tilde{t}_{ij}$ is the effective amplitude for an 
electron to hop from site $i$ to site $j$, and a nearest neighbor pair attraction term with strength
$\tilde{J}_{ij}$. The parameters (i.e.~$\tilde{t}_{ij}$ and $\tilde{J}_{ij}$) of both are strongly affected by
correlations.

In the case of strong local repulsion such as large Mott-Hubbard $U$ ($>>t$) \cite{PFazekas}, one uses the
same form as in its absence, but with renormalized Hamiltonian parameters which are assumed to describe the
entire effect of $U$ (at least at low energies). This implies (for example with renormalized hopping
$\tilde{t}_{ij}$ replacing $t_{ij}$) that there are good mobile quasiparticles, albeit with renormalized
dispersion. The well known single-site Gutzwiller renormalization factor \cite{PFazekas,PWAnderson,BEdegger},
for example, projects out doubly occupied sites completely and is a good approximation for $U\rightarrow \infty$. It
assumes, for small hole density $x$, a multiplicative factor $g_t=2x/(1+x)$ for $t_{ij}$ and $g_s=4/(1+x)^2$
for $J_{ij}$ (i.e. $\tilde{t}_{ij}=g_tt_{ij}$ and $\tilde{J}_{ij}=g_sJ_{ij}$). The bare hopping amplitude $t_{ij}$ 
involves the nearest neighbor ($t$), next-nearest neighbor ($t'$) and further neighbor ($t''$) hopping terms. 
In our calculations, we use the above mentioned homogeneous Gutzwiller approximation with standard values for
$t_{ij}$ ($t=300$ meV, $t'/t=-1/4$ and $t''=0$, see e.g.~\cite{AParamekanti1}). The value of $J_{ij}$ does
not appear explicitly in the calculation, as we elucidate later. The $J_{ij}$ term is the well known
superexchange (e.g.~$J_{ij}\propto t^2_{ij}/U$ in single site Hubbard model \cite{PFazekas}) for
large $U$. We have written the conventional $J_{ij}\mathbf{S}_i.\mathbf{S}_j$ term as a pair attraction
$-J_{ij}b^\dagger_{ij}b_{ij}$ \cite{PWAnderson1,GBaskaran} using the identity
$(\mathbf{S}_i.\mathbf{S}_j-\hat{n}_i\hat{n}_j/4)=-b^\dagger_{ij}b_{ij}$ valid for spin-$\frac{1}{2}$ particles
($\mathbf{S}_i$ and $\hat{n}_i$ are the spin and number operators at site $i$, respectively).
There is thus strong attractive nearest-neighbor spin singlet pairing term in the Hamiltonian, given that the
antiferromagnetic coupling $J_{ij}$ is known to be large ($\sim 1500$ K) \cite{MAKastner} for undoped cuprates ($x=0$).

 The above identity between nearest-neighbor AF Heisenberg spin interaction and spin singlet Cooper pair
attraction means that exact solution of Eq.\eqref{eq.Hamiltonian} with either term would lead to the same
result. Because of the fact that antiferromagnetic LRO disappears for surprisingly small 
hole doping $x$ and is replaced by superconducting order, and that the holes are quite mobile, we use the singlet bond pair 
form in Eq.\eqref{eq.Hamiltonian} as being natural and accurate for good approximations, e.g.~mean field
theory, and assume a homogeneous system as will naturally arise for mobile holes. 

Very generally, the bond-pair self interaction term 
in Eq.\eqref{eq.Hamiltonian} can be written via the exact Hubbard-Stratonovich transformation
\cite{ZTesanovic} as a time (and space) dependent bond pair potential acting on electrons and characterized 
by a field $\psi_m(\tau)$ with a Gaussian probability distribution 
[here $\tau$ is the imaginary time; $0<\tau<\beta=1/(k_BT)$]. The saddle point of the resulting action in the static limit 
gives rise to the conventional mean field 
approximation in which the second term in Eq.\eqref{eq.Hamiltonian} is written as 
\begin{eqnarray}
-\tilde{J}_{ij}(\langle b^\dagger_{ij} \rangle b_{ij}+b^\dagger_{ij}\langle b_{ij}\rangle -\langle b^\dagger_{ij}\rangle \langle b_{ij}\rangle)
\label{eq.MFTDecoupling}
\end{eqnarray}  
and the average $\psi_m=\psi_{ij}\propto \tilde{J}_{ij}\langle b_{ij} \rangle$ is determined self-consistently
(mean field theory).

The effective Hamiltonian we use (see Section \ref{Sec.Methods}) is of the form of Eq.\eqref{eq.Hamiltonian} with the 
second term in it replaced by Eq.\eqref{eq.MFTDecoupling}. This describes two coupled fluids, namely a fermionic fluid and 
a bosonic fluid, represented respectively by the on-site electron field $a^\dagger_{i\sigma}$ and the bond Cooper pair field
$\psi_{ij}=\psi_m$. The properties of $\psi_m$ needed in our calculation are its mean value $\langle \psi_m \rangle$
(nonzero below $T_c$) and the fluctuation part of the correlation function $\langle \psi_m\psi^*_n \rangle$
(whose {\it universal} form for large $|\mathbf{R}_m-\mathbf{R}_n|$ near $T_c$ is what we use). These arise from inter-site
interactions of the Cooper pairs $\psi_m$; our results are independent of the details of these interactions. A
nearest neighbor interaction of the form $C\sum_{<mn>}(\psi^*_m\psi_n+\psi_m\psi^*_n)$, with $C>0$, has been
mentioned earlier as a natural possibility \cite{SBanerjee}. It may arise \cite{SBanerjee} with $C\propto x$ in a strong
correlation picture due to diagonal hole hopping $t'$. 

For a translationally invariant system described by Eq.\eqref{eq.Hamiltonian}, the electron Green's function
satisfies the Dyson equation\cite{GDMahan,AAltland}
\begin{eqnarray}
G^{-1}(\mathbf{k},i\omega_n)&=&(G^0)^{-1}(\mathbf{k},i\omega_n)-\Sigma(\mathbf{k},i\omega_n),
\label{eq.DysonEquation}
\end{eqnarray}
where $\Sigma(\mathbf{k},i\omega_n)$ ($\omega_n=(2n+1)\pi/\beta$ is the fermionic Matsubara frequency with
$n$ as an integer) is the irreducible 
self energy, originating from the coupling between bond pairs and electrons with bare propagator
$G^0(\mathbf{k},i\omega_n)=(i\omega_n-\xi_\mathbf{k})^{-1}$ where $\xi_\mathbf{k}=\epsilon_\mathbf{k}-\mu$ and 
$\epsilon_\mathbf{k}$ is the Fourier transform of the hopping $\tilde{t}_{ij}=\tilde{t}_{\mathbf{i}-\mathbf{j}}$ 
($\mu$ is the chemical potential). 

 We use a well-known bosonic fluctuation exchange
approximation for $\Sigma(\mathbf{k},i\omega_n)$ which captures the leading effect of this coupling beyond the
`Hartree' approximation and is shown diagrammatically in Fig.\ref{fig.BondLattice}{\bf b}. This diagram describes the 
exchange of a Cooper pair fluctuation by an electron. In common with general practice, 
we find $\Sigma$ (and thence $G$) by inserting $G^0$ instead of $G$ in the expression for it 
[see Eq.\eqref{eq.StaticSelfEnergy} below]. This is known to be generally quite accurate \cite{GDMahan}, e.g.~for the 
coupled electron-phonon system.

Close to $T_c$, the temporal decay of long-wavelength fluctuations is specially slow (dynamical critical
slowing down \cite{PMChaikin}) so that they can be regarded as quasistatic \cite{SBanerjee} 
(i.e. the characteristic frequency scale $\omega_0<<k_BT$). Though the decay is `slow', we assume, as is
generally done, that the system is homogeneous. There are \emph{annealing} processes which make it so on the experimental 
time scale. (There may be intrinsic as well as extrinsic sources of static quenched disorder, e.g., due to the very process 
of doping itself \cite{HAlloul}). Even their effect can be included by approximate configuration averaging.
Our approach is similar to that for static critical phenomena where also a homogeneous system is used and all
the fluctuations are thermal. The self energy $\Sigma$ can then be expressed as
\begin{eqnarray}
&&\Sigma(\mathbf{k},i\omega_n)=\nonumber \\
&& -\frac{1}{N}\sum_{\mathbf{q},\mu,\nu} G^0(-{\mathbf{k}+2\mathbf{q}},-i\omega_n)
D_{\mu\nu}(2\mathbf{q})f_\mu(\mathbf{k},\mathbf{q})f_\nu(\mathbf{k},\mathbf{q})\nonumber \\
\label{eq.StaticSelfEnergy}
\end{eqnarray}
where $N$ is the total number of Cu sites on a single $\mathrm{CuO_2}$ plane and $\mu$, $\nu$ refer to the direction of 
the bond i.e. $x$ or $y$. $D_{\mu\nu}(2\mathbf{q})=\sum_\mathbf{R} D_{\mu\nu}(\mathbf{R})
\exp{(-i2\mathbf{q}.\mathbf{R})}$ is the static pair propagator of Cooper pair fluctuations with 
$D_{\mu\nu}(\mathbf{R}_{i\mu}-\mathbf{R}_{j\nu})=\langle \psi_{i\mu}\psi^*_{j\nu}\rangle$. 
The quantity $f_\mu(\mathbf{k},\mathbf{q})=\cos[(k_\mu-q_\mu)a]$ is a form factor arising from the coupling between 
a tight-binding lattice electron and a nearest-neighbour bond pair. Because of the $d$-wave LRO described as
`$\mathrm{Ne'el}$' order\cite{SBanerjee} of the `planar spin' $\psi_m$ in the bipartite bond-centre lattice, the 
standard sublattice transformation (i.e. $\psi_m\rightarrow \tilde{\psi}_m=\Delta_m\exp{(i\tilde{\theta}_m)}$ where
$\tilde{\theta}_m=\theta_m$ for $x$-bonds and $\tilde{\theta}_m=\theta_m+\pi$ for $y$-bonds) implies
$D_{xx}(\mathbf{R})=D_{yy}(\mathbf{R})=-D_{xy}(\mathbf{R})=-D_{yx}(\mathbf{R})\equiv D(\mathbf{R})$ where 
$D(\mathbf{R})$ can be written as
\begin{eqnarray}
D(\mathbf{R}_m-\mathbf{R}_n)=\langle \tilde{\psi}_m\rangle \langle
\tilde{\psi}^*_n\rangle+\widetilde{D}(\mathbf{R}_m-\mathbf{R}_n)\,.
\label{eq.UrsellFunction}
\end{eqnarray}
Here $\widetilde{D}(\mathbf{R})$ is the fluctuation term. The LRO part $\langle \tilde{\psi}_m\rangle\equiv \Delta_d$ leads to a 
$d$-wave Gor'kov like gap with $\Delta_\mathbf{k}= (\Delta_d/2) (\cos{k_xa}-\cos{k_ya})$; the corresponding electron self 
energy is $\Sigma(\mathbf{k},i\omega_n)=\Delta_\mathbf{k}^2/(i\omega_n+\xi_\mathbf{k})$. In widely used 
phenomenological analyses \cite{MRNorman2} 
of ARPES data, this form is used above $T_c$ with lifetime effects, both diagonal and off-diagonal 
in particle number space, added to $\Sigma$, i.e.~
\begin{eqnarray}
\Sigma(\mathbf{k},\omega)&=&-i\Gamma_1+\frac{\Delta_\mathbf{k}^2}{\omega+\xi_\mathbf{k}+i\Gamma_0}
\end{eqnarray} 
Here $\Gamma_1$ is single-particle scattering rate and $\Gamma_0$ is assumed to originate due
to finite life-time of \emph{preformed} $d$-wave pairs \cite{MRNorman2}.

We propose here that as described above, the electrons move (above $T_c$) \emph{not} in a pair field with $d$-wave LRO which
decays in time at a rate put in by hand, but in a nearly static pair field with growing
correlation length $\xi$.  We assume, as appears quite natural for a system with characteristic length scale
$\xi$ that $\widetilde{D}(\mathbf{R})\sim \exp(-R/\xi)$ for large $R$, while $\xi$ diverges at $T_c$. 
This natural form for $\widetilde{D}(R)$ is found, for example, in the Berezinskii-Kosterlitz-Thouless (BKT) theory 
\cite{VLBerezinskii,JMKosterlitz1,JMKosterlitz2,PMChaikin} for two dimensions and in a Ginzburg-Landau (GL) theory for all dimensions. As described 
in Appendix \ref{App.FermiArc}, it so happens that for large correlation lengths, $\Sigma(\mathbf{k},i\omega_n)$ is nearly 
the same as that for preformed $d$-wave symmetry pairs. Below $T_c$, $\widetilde{D}(R)$ decays as a power law (i.e.
$\widetilde{D}(R)\sim R^{-\eta}$ with $\eta>0$) due to order parameter phase or `spin wave'-like fluctuations 
(see the discussion in Section \ref{Sec.Methods}). We find here the consequences of these for the spectral 
function, i.e.~
\begin{eqnarray} 
A(\mathbf{k},\omega)&\equiv&-\frac{2}{\pi}\mathrm{Im}\left[G(\mathbf{k},i\omega_n\rightarrow\omega+i\delta)\right],
\label{eq.SpectralFunction}
\end{eqnarray}
 measured in ARPES \cite{ADamascelli}. $G(\mathbf{k},i\omega_n)$ is obtained from
Eq.\eqref{eq.DysonEquation} with the self energy calculated from Eq.\eqref{eq.StaticSelfEnergy}
using the aforementioned forms of the pair propagator $\widetilde{D}(\mathbf{R})$ [or ${D}(\mathbf{R})$] for
temperatures above and below $T_c$.

\section{Electron self energy and phenomenological inputs} \label{Sec.Methods}
The tight binding lattice Hamiltonian $\mathcal{H}$ is given by Eq.\eqref{eq.Hamiltonian}. Near $T_c$ the time 
dependence of $\psi_{ij}$ can be neglected (except when $T_c$ itself is close to zero where time dependence of
$\psi_{ij}$ has significant effects) so that this becomes
\begin{eqnarray}
\mathcal{H}&=&-\sum_{ij,\sigma} \tilde{t}_{ij}a^\dagger_{i\sigma} a_{i\sigma}-\sum_{<ij>}
(\psi_{ij}b^\dagger_{ij}+\mathrm{h.c})\,. \label{eq.MFTHamiltonian}
\end{eqnarray}
Eq.\eqref{eq.MFTHamiltonian} describes electrons moving in a static but (in general) spatially fluctuating pair potential 
$\psi_{ij}$ whose correlation length (in our case) diverges as $T\rightarrow T_c$. The effect of the long distance 
fluctuations on the electrons is captured by the self energy shown in Fig.\ref{fig.BondLattice}{\bf b} and algebraically 
described by Eq.\eqref{eq.StaticSelfEnergy}, with the significant fluctuation wave vector $q\sim \xi^{-1}<< a^{-1}$. In a 
regime where the fluctuations in the real pair magnitude $\Delta_m$ are short ranged,
\begin{eqnarray}
D(\mathbf{R})\simeq \langle \Delta(\mathbf{R})\rangle \langle \Delta(\mathbf{0})\rangle
\langle e^{i[\tilde{\theta}(\mathbf{R})-\tilde{\theta}(\mathbf{0})]}
\rangle\equiv
\bar{\Delta}^2 \bar{D}(R) \nonumber \\\label{eq.PairCorrelator}
\end{eqnarray} 
for large $R$ ($R>>a$) (as evident from Eq.\eqref{eq.UrsellFunction}, $\widetilde{D}(R)\simeq
\bar{\Delta}^2(\bar{D}(R)-|<e^{i\tilde{\theta}(\mathbf{0})}>|^2)$). This 
decoupling between magnitude and long distance phase correlations is accurate for $x\aplt x_\mathrm{opt}$, a
manifestation of which is the separation between 
$T^*$ and $T_c$. Our calculations based on Eq.\eqref{eq.PairCorrelator} are
therefore reliable in this doping range. We use the general form 
\cite{PMChaikin} $\bar{D}(R)=(\tilde{\Lambda}R)^{-\eta}\exp{(-R/\xi)}$ (with $\tilde{\Lambda}\sim a^{-1}$) in
Eq.\eqref{eq.StaticSelfEnergy}, expanding $\xi_{\mathbf{k}-2\mathbf{q}}$ and
$f_\mu(\mathbf{k},\mathbf{q})$ in powers of $\mathbf{q}$ for $qa<<1$. The self
energy $\Sigma(\mathbf{k},i\omega_n)$ (Eq.\eqref{eq.StaticSelfEnergy}) is (see
Appendix \ref{App.SelfEnergy}) then
\begin{eqnarray}
\Sigma(\mathbf{k},i\omega_n)&\simeq&\frac{-i~\mathrm{sgn}(\omega_n)\Gamma(1-\eta)\bar{\Delta}_\mathbf{k}^2}{\left(\tilde{\Lambda}
\mathrm{v}_\mathbf{k}\right)^\eta
\left(v_\mathbf{k}/\xi-i~\mathrm{sgn}(\omega_n)(i\omega_n+\xi_\mathbf{k})\right)^{1-\eta}}\,. \nonumber \\
&&\label{eq.MatsubaraSelfEnergy}
\end{eqnarray}
In this equation, $\Gamma$ is the well known gamma function and
$\mathbf{v}_\mathbf{k}=\frac{1}{a}\frac{\partial\xi_\mathbf{k}}{\partial\mathbf{k}}$ (with
$v_\mathbf{k}=|\mathbf{v}_\mathbf{k}|$) is the velocity (expressed
in units of energy) obtainable from the energy dispersion $\xi_\mathbf{k}$. There is a relatively small particle-hole
asymmetric part that adds to $\bar{\Delta}_\mathbf{k}^2$ (in the numerator of
Eq.\eqref{eq.MatsubaraSelfEnergy}) a term proportional to $(i\omega_n+\xi_\mathbf{k})$,
and is ignored henceforth. The above self energy does not affect the nodal quasiparticles owing to the
$\mathbf{k}$ dependence of $\bar{\Delta}_\mathbf{k}=(\bar{\Delta}/2)(\cos{k_xa}-\cos{k_ya})$ (of which the
second term arises from the form factor appearing in Eq.\eqref{eq.StaticSelfEnergy}). Here it is very
important to mention that the quantity $\bar{\Delta}_\mathbf{k}$ should \emph{not} be confused with the spectral gap
$\Delta_\mathbf{k}$ that we are going to define below from the position of peak (as a function of $\omega$ at
a fixed $\mathbf{k}$) of the spectral function $A(\mathbf{k},\omega)$. The spectral function $A(\mathbf{k},\omega)$
(Eq.\eqref{eq.SpectralFunction}) is calculated using the self energy of Eq.\eqref{eq.MatsubaraSelfEnergy}, and
the consequent gap ($\Delta_\mathbf{k}$) in the electronic spectra is in general different (except at $T=0$), due to pair 
fluctuations, from the \emph{input} $\bar{\Delta}_\mathbf{k}$ (or $\bar{\Delta}$ that appears in
Eq.\eqref{eq.PairCorrelator}). An additional term
describing the coupling of electrons to short range phase fluctuations (present in our theory, but not
discussed here) is probably relevant for nodal quasiparticles. The self energy of Eq.\eqref{eq.MatsubaraSelfEnergy} has a 
nonvanishing imaginary part as $\omega\rightarrow 0$ because of the decay of an electron into one of opposite
momentum and small $\mathbf{q}$ Cooper pair fluctuations (see Fig.\ref{fig.BondLattice} {\bf b}). Thus the electronic
system is a non Fermi liquid.

\begin{figure} [hbt!]
\begin{center}
\begin{tabular}{c}
\includegraphics[width=9cm]{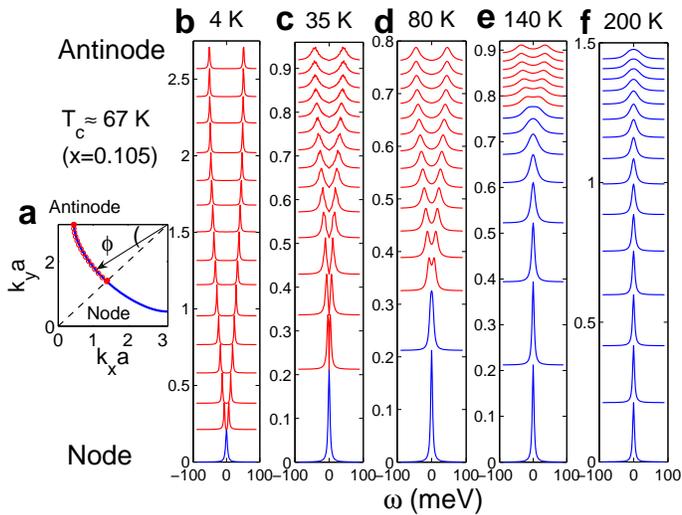}
\end{tabular}
\end{center}
\caption{{\bf The Fermi surface and the spectral density along it.} {\bf a}, Fermi surface (FS) in the first
quadrant of the Brillouin zone for $x\simeq 0.11$, a typical value. The angular position $\phi$ on the Fermi surface is 
shown. The antinode corresponds to $\phi=0^\circ$ and node to $\phi=45^\circ$. {\bf b-f}, Single-particle spectral density $A(\mathbf{k},\omega)$ at 15 equally spaced points (red circles) on the 
Fermi surface in {\bf a} shown vertically shifted from node (bottom) to antinode (top), as a function of
temperature at $x\simeq 0.11$. Blue and red lines correspond to Fermi arc and gapped portions of FS, respectively.}
\label{fig.EDC}
\end{figure}

In order to plot the spectral peaks, e.g.~in Fig.\ref{fig.EDC}{\bf a}, we use a small lifetime broadening
($\sim 3$ meV, which is much less than the typical instrumental resolution of $\sim 10$ meV of ARPES
\cite{ADamascelli,JCCampuzano1}). The nodal peaks, which would be $\delta$-functions in its absence, can
then be detected clearly.

 Above $T_c$, we use in Eq.\eqref{eq.MatsubaraSelfEnergy} the BKT form 
\begin{eqnarray}
\xi(x,T)&\simeq& a\exp{[b'(x)/\sqrt{T/T_c(x)-1}]} \label{eq.BKTCorrelationLength}
\end{eqnarray}
 and $\eta=\eta_c\equiv T_c/(2\pi\rho_s(T_c))=0.25$,
corresponding to the universal Nelson-Kosterlitz jump value\cite{PMChaikin}, the BKT transition temperature
(say $T_\mathrm{BKT}$) has been taken to be actual observed \cite{MRPersland} $T_c$ for the cuprates which
have small interlayer coupling. The observed $T_c$ can be substantially higher ($\sim 5-10$ K) than the underlying 
BKT transition temperature. This temperature $T_\mathrm{BKT}$, though a \emph{fictitious} transition temperature that can 
only be observed by \emph{switching off} interlayer coupling between $\mathrm{CuO}_2$ planes, is expected to
control the correlation length above $T_c$, away from a narrow 3D-XY critical region \cite{GBlatter}. The use
of experimentally measured $T_c$ in Eq.\eqref{eq.BKTCorrelationLength} as well as the precise value of the exponent $b'(x)$ 
are not crucial for the present calculations, although certain details, e.g. the $T$-dependence of the Fermi arc length 
close to $T_c$ (within $\sim 10$ K), can be sensitive to these. For the 2D-XY model \cite{POlsson} $b'\simeq 1.6$. No 
systematic estimate of either $\xi(x,T)$ or $b'(x)$ exists for the cuprate superconductors. Hence we work 
with $b'(x)$ obtained using the phenomenological functional of Ref.\cite{SBanerjee} (see Appendix
\ref{App.CorrelationLength} for details). The Kosterlitz RG equations \cite{JMKosterlitz1} predict the asymptotic BKT form 
for $\xi(T)$ to be valid only over a narrow critical regime above $T_c$. However, in practice, the BKT form is 
found \cite{POlsson} to fit the Monte Carlo data for $\xi(T)$ for a 2D-XY model very well over a rather large region of 
temperature above $T_c$, till $(\xi/a)\simeq 2$. In the same spirit we use the formula of
Eq.\eqref{eq.BKTCorrelationLength} over a wide range.

Below $T_c$, we have used the self energy of Eq.\eqref{eq.MatsubaraSelfEnergy} which has been evaluated
using the long-distance power-law form $\bar{D}(R)\sim R^{-\eta}$ appropriate for a quasi-LRO state in 2D with
$\eta=(T/2\pi\rho_s)$, as well
as using a $\bar{D}(R)$ in which a small interlayer coupling is incorporated through a simple anisotropic 3D `spin wave' 
approximation (see Appendix \ref{App.PairCorrelator}). Both approximations produce similar results for 
$A(\mathbf{k},\omega)$. The quantity $\eta(x,T)=T/(2\pi\rho_s(x,T))$, with $\rho_s(x,T)$ being the superfluid
density, is an input to both of these and we use 
$\rho_s(x,T)$ estimated from penetration depth data\cite{WAnukool} for Bi2212. In addition, the anisotropic 3D form for 
$\bar{D}(R)$ contains \emph{explicitly} the interlayer coupling characterized by the $c$-axis superfluid density $\rho_s^c$ 
or the anisotropy ratio $\varepsilon=(\rho_s^c/\rho_s)^{1/2}$ ($\sim 10^{-2}$ for Bi2212\cite{TSchneider}).
Close to but below $T_c$, 
the experimental values for $\eta$ exceed (see Fig.\ref{fig.PeakStrength} {\bf a}) by a large amount the upper limit of 
$\eta$ for a pure 2D system i.e. $\eta_c=0.25$. This fact demonstrates the essential role played by the interlayer coupling 
in these systems (the observed LRO below $T_c$ is because of it!). Also, the 2D-BKT results that we use above $T_c$, are not
reliable, due to the importance of 3D fluctuations for a narrow temperature regime close to $T_c$ such that
$\xi(T)>\varepsilon^{-1}c$ ($c$ is the interlayer spacing); we compare our results with ARPES data outside this range, 
i.e.~outside the critical regime. 

Here, it should be
mentioned that we have neglected Coulomb interaction between Cooper pairs, which are charged objects, and
associated quantum phase fluctuation effects, which can be specially prominent near hole densities (extreme
overdoped and underdoped) where $T_c$ vanishes. The power law form  used below $T_c$ in our calculation arises
due to \emph{longitudinal} phase fluctuations and Coulomb interaction can push these up to plasma frequency, which is
quite large ($\sim 1$ eV for cuprates \cite{AParamekanti,LBenfatto}). In this case, the longitudinal phase
fluctuations behave classically and are relevant only above a quantum to classical crossover scale
$T_\mathrm{cl}$. In the absence of dissipation $T_\mathrm{cl}$ has been estimated to be around 100 K, although
dissipation can reduce $T_\mathrm{cl}$ to a much lower value, $\sim$ 20 K, as has been inferred in
Ref.\cite{LBenfatto}. In our calculation, below $T_c$, the effects (e.g.~the deviation of the
$\mathbf{k}$-space gap structure from $d$-wave form) of classical phase fluctuations in electron
spectral function are considerable only for much higher temperatures, typically close to $T_c$ where
$T>T_\mathrm{cl}$. Hence we expect the calculated effects to be genuine and natural consequences for phase fluctuating
superconductors such as cuprates, and not to be changed much by quantum fluctuation effects except near the end
points of $T_c(x)$ curve.  

The other phenomenological input, namely $\bar{\Delta}(x,T)$, has been estimated from ARPES
data\cite{AKanigel1,AKanigel2,JCCampuzano} in the following way. (More details are given in Appendix
\ref{App.FermiArc}). Above $T_c$, the self energy of Eq.\eqref{eq.MatsubaraSelfEnergy} can be written a
simplified form for $\eta=0$ (in the actual calculation, above $T_c$ we use $\eta=\eta_c=0.25$, the BKT jump
value), i.e.
\begin{eqnarray}
\Sigma(\mathbf{k},\omega)&=&\frac{\bar{\Delta}_\mathbf{k}^2}{\omega+\xi_\mathbf{k}+i(v_\mathbf{k}/\xi)}
\label{eq.SelfEnergy_eta0}
\end{eqnarray}
The resulting peak positions of the spectral function $A(\mathbf{k},\omega)$ on the Fermi surface
(i.e.~$\xi_\mathbf{k}=0$) are at $\Delta_\mathbf{k}$ which can be easily obtained as
\begin{subequations}
\begin{eqnarray}
\Delta_\mathbf{k}&=&0~~~~~~~~~~~~~~~~~~~~~~\mathrm{for}~~~|\bar{\Delta}_\mathbf{k}|\leq
\frac{v_\mathbf{k}}{\sqrt{2}\xi} \\
\Delta_\mathbf{k}^2&=&
\bar{\Delta}_\mathbf{k}^2-\left(\frac{v_\mathbf{k}}{\sqrt{2}\xi}\right)^2~~~\mathrm{otherwise}
\end{eqnarray} 
\end{subequations}
The above expressions imply that, above $T_c$, there is a \emph{gapless} (i.e.~spectral peak at zero energy)
portion corresponding to the $\mathbf{k}$ values such that 
\begin{eqnarray}
|\bar{\Delta}_\mathbf{k}(T)|&\leq& \frac{v_\mathbf{k}}{\sqrt{2}\xi(T)} \label{eq.FermiArcCriterion}
\end{eqnarray}
 As evident from this relation, the extent of this gapless portion along the Fermi surface 
is temperature dependent since both $\bar{\Delta}_\mathbf{k}$ and $\xi$ depend on temperature. This is the
phenomena of `Fermi arc' \cite{MRNorman1}; we discuss it in more detail in the next section
(Section \ref{Sec.Results}). This criterion (Eq.\eqref{eq.FermiArcCriterion}) enables us to deduce a temperature,
$T_\mathrm{an}$, at which the antinodal gap,
$\Delta_\mathrm{an}=\sqrt{\bar{\Delta}_\mathrm{an}^2-(v_\mathrm{an}/(\sqrt{2}\xi))^2}$, gets completely filled
in, i.e.~$\Delta_\mathrm{an}(T_\mathrm{an})=0$ so that the tip of the Fermi arc reaches the antinodal Fermi
momentum $\mathbf{k}_\mathrm{an}$ (here $\bar{\Delta}_\mathrm{an}\equiv \bar{\Delta}_{\mathbf{k}_\mathrm{an}}$
and $v_\mathrm{an}\equiv v_{\mathbf{k}_\mathrm{an}}$). This implies a self-consistent condition for
$T_\mathrm{an}$
\begin{eqnarray}
\bar{\Delta}(x,T_\mathrm{an})&\simeq&
|\bar{\Delta}_\mathrm{an}(x,T_\mathrm{an})|=\frac{v_\mathrm{an}(x)}{\sqrt{2}\xi(x,T_\mathrm{an})}
\label{eq.PseudogapFilling}
\end{eqnarray}
With the aid of the above, we estimate the phenomenological input $\bar{\Delta}(x,T)$ at the ARPES-measured
pseudogap temperature $T^*$ \cite{JCCampuzano,AKanigel2} by identifying it with $T_\mathrm{an}$,
i.e.~$\bar{\Delta}(T^*)\simeq v_\mathrm{an}/(\sqrt{2}\xi(T^*))$ (the right hand side of this relation is
already known as we have estimated $\xi$ from Eq.\eqref{eq.BKTCorrelationLength} and $v_\mathbf{k}$ is
obtained from the band-structure $\xi_\mathbf{k}$). Also at $T=0$, $\bar{\Delta}(0)$ can be
deduced from the zero temperature antinodal gap $\Delta_\mathrm{an}(0)$ measured in ARPES \cite{JCCampuzano},
as there $\Delta_\mathrm{an}(0)=(\bar{\Delta}(0)/2)(\cos(k_{\mathrm{an},x}a)-\cos(k_{\mathrm{an},y}))$. We use a simple
interpolation formula, $\bar{\Delta}(x,T)=\bar{\Delta}(x,0)[1-\sinh(\alpha(x)T)]$ and utilize the knowledge of 
$\bar{\Delta}(T)$ at these two temperatures, namely at $T=0$ and at $T=T^*$, to determine the parameter
$\alpha(x)$ (see Appendix \ref{App.FermiArc}). This chosen form for $\bar{\Delta}(x,T)$ implies that
$\Delta_\mathrm{an}(T)$, for a particular $x<x_\mathrm{opt}$, varies substantially with $T$ only near
$T=T^*\simeq T_\mathrm{an}$. This choice qualitatively mimics the experimentally observed
temperature-dependence \cite{MRNorman1,AKanigel2,WSLee} of the antinodal gap (see Appendix \ref{App.FermiArc}). 

\section{Results} \label{Sec.Results}

\begin{figure*} [hbt]
\begin{center}
\begin{tabular}{ccc}
\includegraphics[height=6cm]{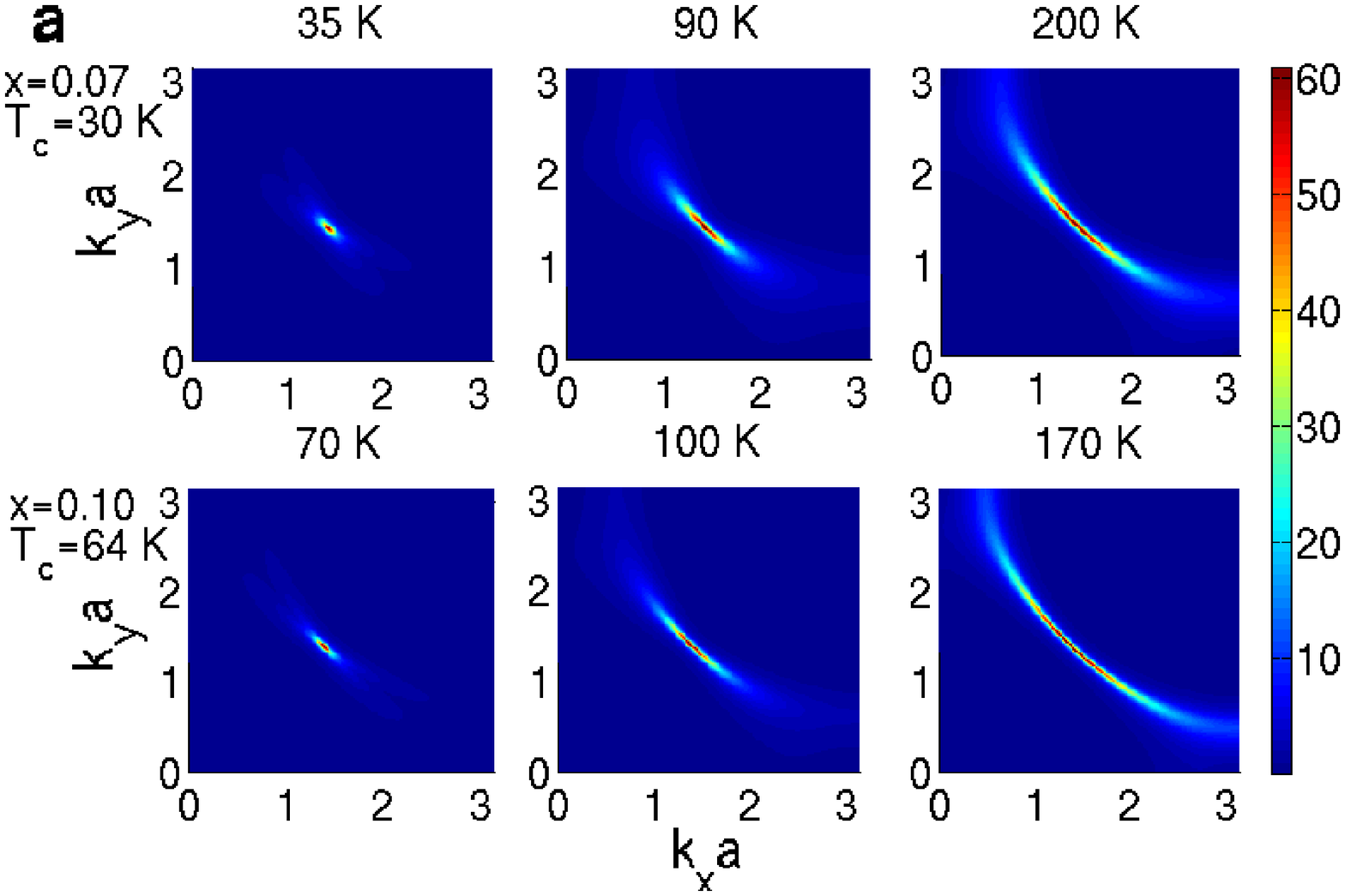}&
\includegraphics[height=6cm]{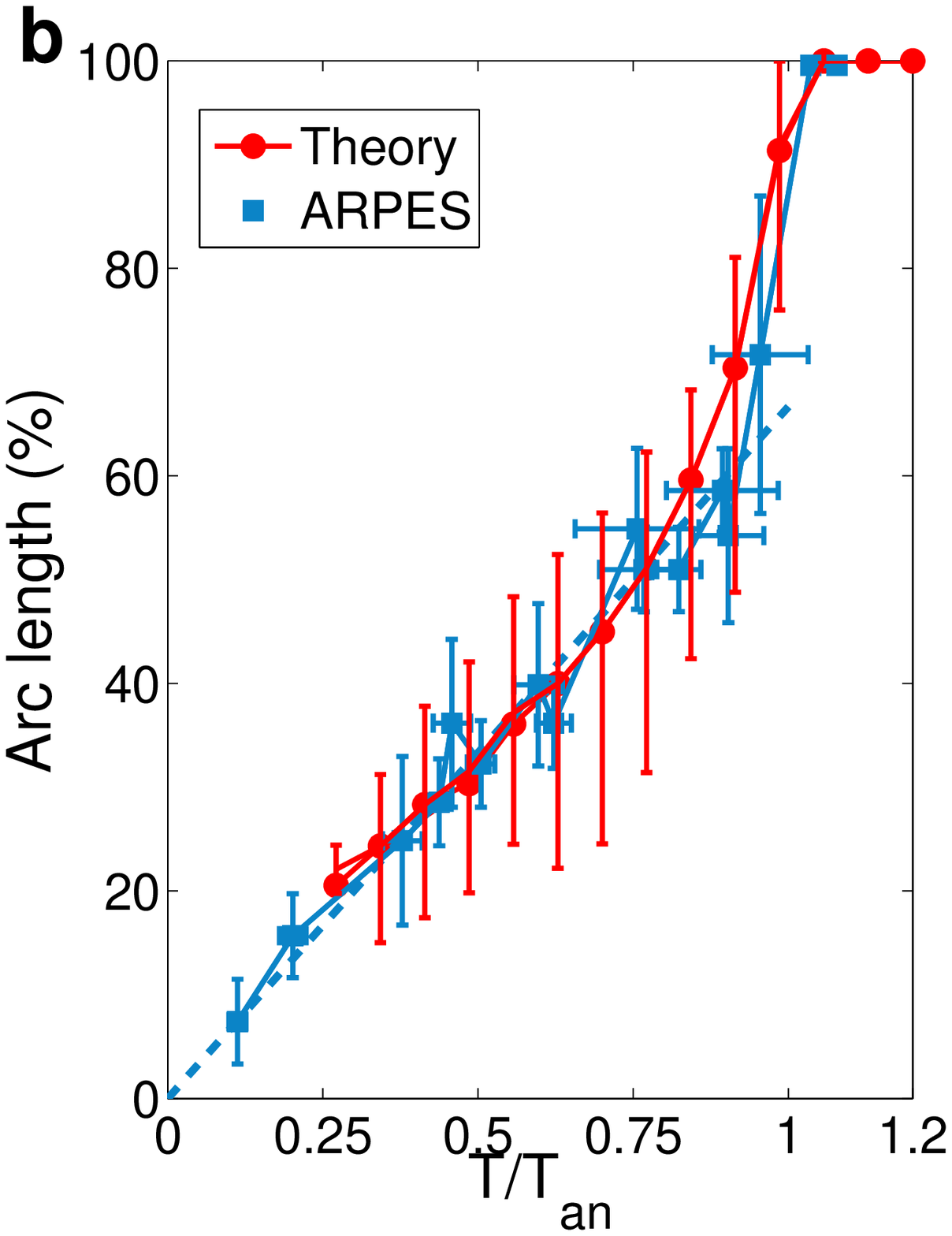}&
\includegraphics[height=6cm]{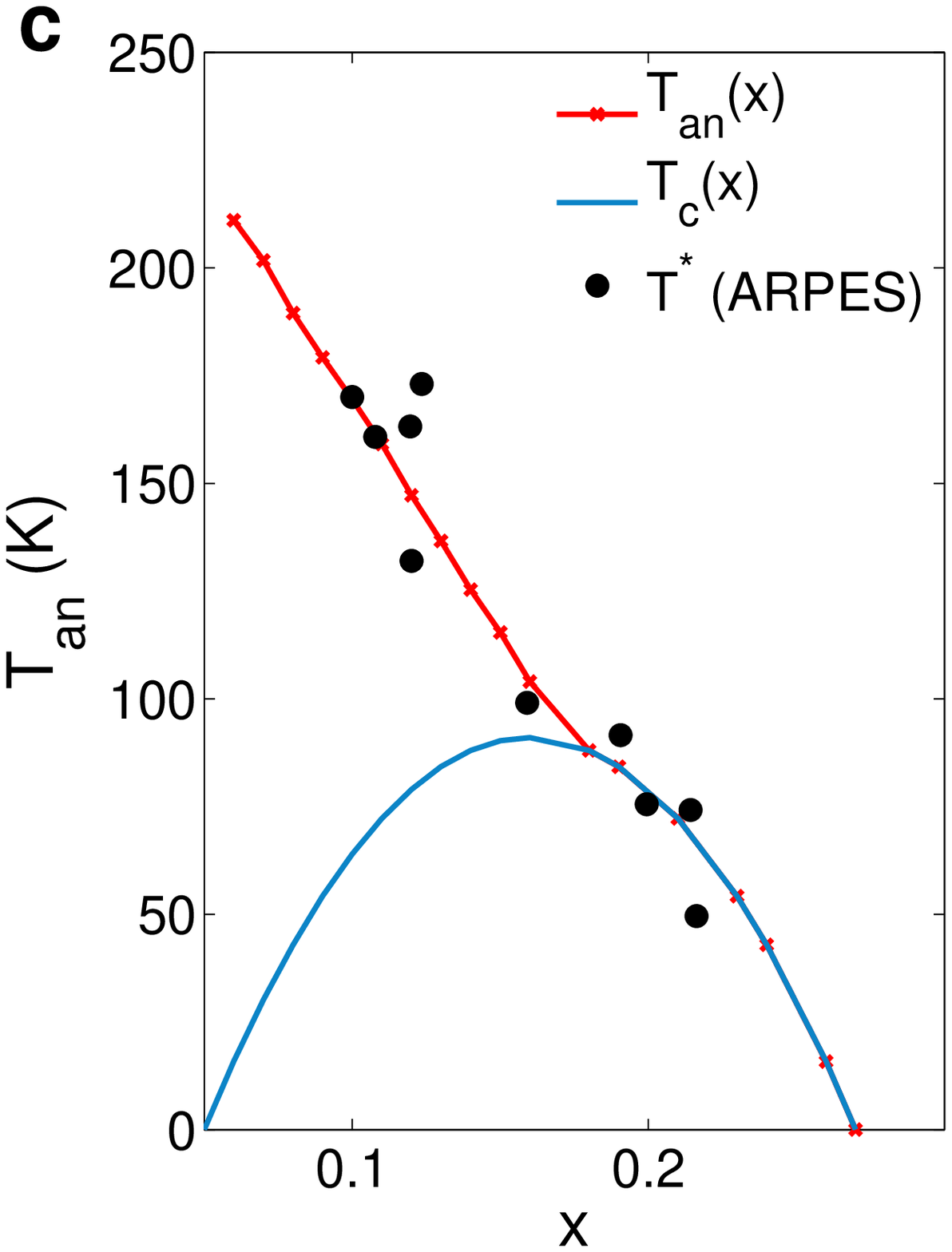}
\end{tabular}
\end{center}
\caption{{\bf Spectral properties above $T_c$.} {\bf a}, Colourmap of
$A(\mathbf{k},\omega=0)$ over the first quadrant of the Brillouin zone for two values of $x$ at
different temperatures. The Fermi arc is easily picked out visually. The colourbar indicates the value of 
$A(\mathbf{k},\omega=0)$ (in units of $3\times 10^{-3}$ $\mathrm{meV}^{-1}$). {\bf b}, The 
arc length vs. $T/T_\mathrm{an}$ curve excluding the temperature region within $\sim 15$ K of $T_c$ (see
Appendix \ref{App.CorrelationLength}), averaged over the 
entire doping range ($x\aplt x_\mathrm{opt}$). On the $y$-axis, $0\%$ is the node and $100\%$ is the antinode. The 
experimental data are from Ref.\cite{AKanigel1}. The vertical error bars in the
theoretical points indicate the variation of the arc length at different $x$ for a given $T/T_\mathrm{an}(x)$.
{\bf c}, The antinodal pseudogap filling temperature $T_\mathrm{an}$ as a function of $x$ is compared
with the data for the pseudogap temperature $T^*$ from ARPES \cite{JCCampuzano,AKanigel2}. The parabolic $T_c(x)$
curve \cite{MRPersland} used in our calculation is also shown.}
\label{fig.FermiArc}
\end{figure*}

The results for spectral density are described below and are compared with ARPES experiments for $\omega \aplt 100$ meV in 
$\mathrm{Bi_2Sr_2CaCu_2O_{8+\delta}}$
(Bi2212) (with maximum $T_c$, $T_c^\mathrm{max}\simeq 92$ K occurring at $x_\mathrm{opt}\simeq0.16$) 
across the underlying Fermi surface (FS) (see Appendix \ref{App.FermiSurface}). As mentioned in the preceding
section, pairing fluctuation related quantities such as $\eta(x,T)$ [$\eta= (T/(2\pi\rho_s))$ below $T_c$ where 
$\rho_s(x,T)$ is the measured superfluid density], $\xi(x,T)$ as well as the average local gap \cite{SBanerjee} 
$\bar{\Delta}(x,T)=\langle\Delta_m\rangle$ act as inputs to our calculation. The GL-like
theory of Ref.\cite{SBanerjee} provides a unified approach in which all the above quantities and many other
results emerge from a 
single assumed free energy functional; this also leads to results \cite{SBanerjee} similar to those described below. As 
remarked earlier, our results are not reliable for extreme underdoped and overdoped regions where
$T_c\rightarrow 0$ and consequently quantum phase fluctuation 
effects can be crucial \cite{SBanerjee,ZTesanovic}.

Our results for the (particle-hole symmetrized) single electron spectral density $A(\mathbf{k},\omega)$ are shown in
Fig.\ref{fig.EDC}{\bf b}-{\bf f}. From the spectral density curves (Fig.\ref{fig.EDC}{\bf b}-{\bf f}) for an 
underdoped system ($x\simeq 0.11$ and $T_c\simeq 67$ K) it is clear that above $T_c$, for some region of the Fermi surface 
centered around the nodal point ($\phi=0$ or $k_x=k_y$), $A(\mathbf{k},\omega)$ peaks at $\omega=0$. The
extent in $\mathbf{k}$ space over the Fermi surface of this 
(ungapped quasiparticle excitation) part, normally identified with the `Fermi
arc'\cite{MRNorman1}, depends on temperature. For any $\mathbf{k}$ outside this region, the
spectral function has a peak at a
nonzero energy; the value of the peak position is identified conventionally with the energy gap
$\Delta_\mathbf{k}$. \emph{The Fermi arc emerges above $T_c$ due to the existence of a finite correlation length in the 
system}, as explained below. 

As discussed in the preceding section (Section \ref{Sec.Methods}), the
self energy expression (Eq.\eqref{eq.MatsubaraSelfEnergy}) obtained by us implies that the gapped portion terminates 
(and the Fermi arc begins) when $|\bar{\Delta}_\mathbf{k}|\leq v_\mathbf{k}/\xi$. One can understand this
relation from the fact that a quasiparticle with momentum $\mathbf{k}$ near the Fermi energy moving with a velocity
$\mathbf{v}_\mathbf{k}$ in a spatially fluctuating superconductor having correlated patches of length
$\xi$ undergoes scattering over a time interval $\tau \apgt \xi/\mathrm{v}_\mathbf{k}$. The energy of
such a quasiparticle is $\sim \bar{\Delta}_\mathbf{k}$ so that from the uncertainty relation $\tau
\bar{\Delta}_\mathbf{k}\apgt 1$ or $(\xi/v_\mathbf{k})\bar{\Delta}_\mathbf{k}\apgt 1$. Above $T_c$, even in the 
region of the Fermi surface where $\Delta_\mathbf{k}\neq 0$, there is (both in our calculations and in experiments)
substantial spectral density at zero excitation energy. The simultaneous presence of a nonzero $\Delta_\mathbf{k}$ 
and finite spectral density at zero frequency is the operational definition of pseudogap. 
Below $T_c$, the calculated spectral density has two symmetrical peaks at nonzero
$\Delta_\mathbf{k}$ for all $\mathbf{k}$ except at the node, where the peak is at zero energy.

Fig.\ref{fig.FermiArc}{\bf a} gives an overall pictorial view of the development of the Fermi arc as a function of $x$
and $T$ above $T_c$. In Fig.\ref{fig.FermiArc}{\bf b}, we plot the arc length versus the reduced temperature,
namely $(T/T_\mathrm{an})$, where $T_\mathrm{an}$ is the temperature at which antinodal pseudogap fills up. The
nearly straight line obtained by us over a large range of doping ($0.07\leq x \leq 0.15$) is seen to compare
well with ARPES data over the same range \cite{AKanigel1}. Qualitatively, since the $\mathbf{k}$ points on
the Fermi arc  satisfy the condition $|\bar{\Delta}_\mathbf{k}|\aplt \mathrm{v}_\mathbf{k}/\xi$,
this condition is met for a larger $\mathbf{k}$ as $\xi(T)$ decreases on increasing $T$. Thus the arc
length increases with increasing $T$, 
and at $T_\mathrm{an}(x)\sim T^*(x)$ it is as large as the entire Fermi surface. Conversely,
it shrinks to zero at $T_c$ and below because $\xi\rightarrow \infty$ there.

\begin{figure}[hbt]
\begin{center}
\includegraphics[height=7cm]{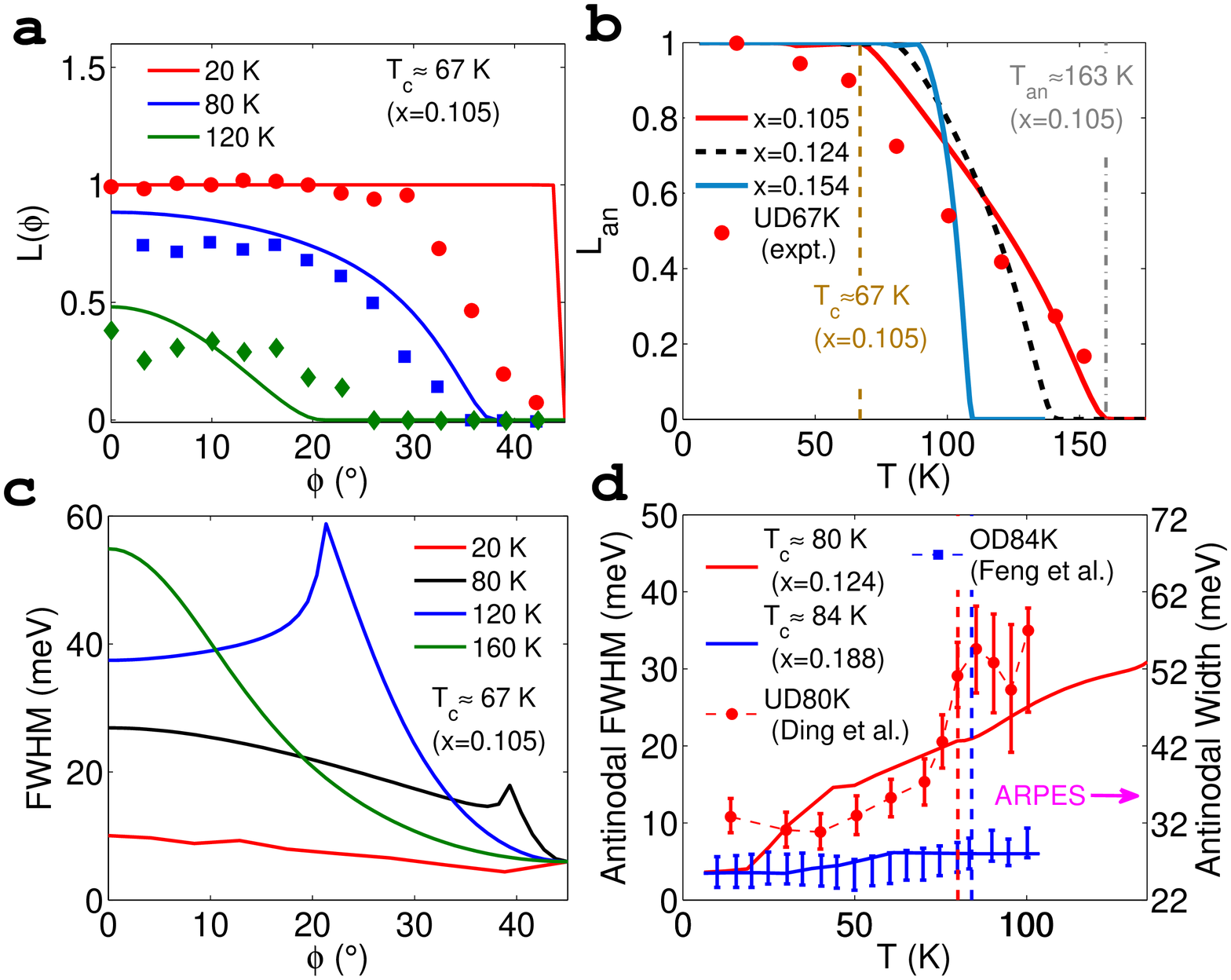}
\end{center}
\caption{{\bf Spectral properties both above and below $T_c$.} {\bf a}, Loss of low energy spectral weight $L(\phi)$ 
(see text for definition) for three temperatures at $x\simeq 0.11$. The symbols correspond to the experimental 
data \cite{AKanigel2} at same temperatures for an underdoped (UD) sample with $T_c=67$ K, $T^*\simeq 150$ K. {\bf b}, Filling of
the antinodal pseudogap (quantified by $L_\mathrm{an}$) with temperature for three values of $x$. {\bf c}, Width (FWHM) of 
the spectral peaks along the FS for $x\simeq 0.11$. {\bf d}, Variation of the width of the antinodal peak with
temperature ($<T_\mathrm{an}$) for an underdoped ($x\simeq0.12$) and an overdoped ($x\simeq 0.19$) sample, compared with 
antinodal widths at the same $x$ values extracted from ARPES data in Refs.\cite{HDing,DLFeng}. Vertical dashed lines
indicate the corresponding $T_c$'s.}
\label{fig.SpectralFeatures}
\end{figure}

\begin{figure}[hbt]
\begin{center}
\includegraphics[height=6cm]{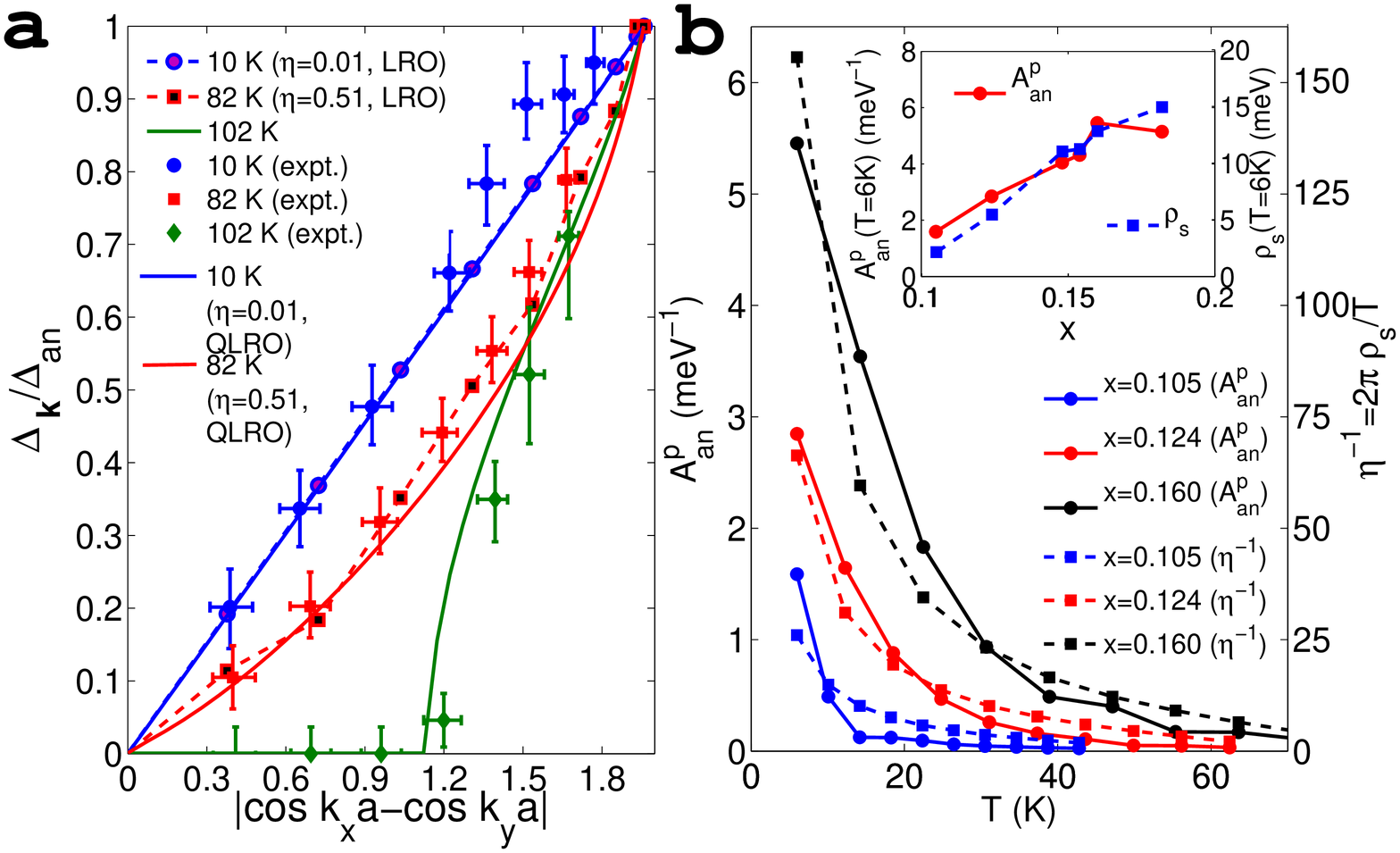}
\end{center}
\caption{{\bf Spectral properties below $T_c$.} {\bf a}, Deviation of $\Delta_\mathbf{k}$ from the $d$-wave form,
$|\cos{k_xa}-\cos{k_ya}|$, below $T_c$ for $x=0.16$ is compared with $\Delta_\mathbf{k}$ obtained in ARPES\cite{WSLee}. 
At low temperature ($T=10$ K), $\Delta_\mathbf{k}$ follows the canonical $d$-wave form (a straight line), but bends away 
considerably from it close to $T_c$. For $T<T_c$, we have shown $\Delta_\mathbf{k}$
obtained for both the true LRO phase (i.e.~$D(R\rightarrow \infty)=|\langle\psi_m\rangle|^2\neq 0$) and the quasi-LRO or 
QLRO phase (i.e.~$D(R)\sim R^{-\eta}\xrightarrow{R\rightarrow \infty}0$) (see Methods). The quantity
$\eta\propto T/\rho_s$ or the superfluid density $\rho_s(x,T)$ has been estimated
from the penetration depth data of Ref.\cite{WAnukool}. {\bf b}, The antinodal peak height
$A^p_\mathrm{an}$ vs.~$T$ tracks $\rho_s(T)/T$ vs.~$T$. {\bf Inset} At low temperatures ($T=6$ K) $A^p_\mathrm{an}(x)$ 
closely follows $\rho_s(x)$, especially in the underdoped side ($x\aplt 0.16$).}
\label{fig.PeakStrength}
\end{figure}

Fig.\ref{fig.FermiArc}{\bf c} shows $T_\mathrm{an}(x)$ together with $T_c(x)$ \cite{MRPersland} and experimental data 
for $T^*(x)$ \cite{JCCampuzano} for Bi2212. As mentioned above, $T_\mathrm{an}$ is taken to coincide with the temperature at 
which the Fermi arc takes up the full FS,
i.e.~$|\bar{\Delta}_\mathrm{an}(x,T_\mathrm{an})|=\mathrm{v}_\mathrm{an}/(\sqrt{2}\xi(x,T_\mathrm{an}))$. 
Hence $T_\mathrm{an}(x)$ is decided by
$\bar{\Delta}(x,T)\sim |\bar{\Delta}_\mathrm{an}(x,T)|$ and $\xi(x,T)$, both of which are
phenomenological inputs to our calculations ($\mathrm{v}_\mathrm{an}$ and
$\mathbf{k}_\mathrm{an}$ are obtained from the energy dispersion $\xi_\mathbf{k}$). For $\xi(x,T)$ we use the
BKT form of Eq.\eqref{eq.BKTCorrelationLength}. We have already discussed the choice of the exponent $b'(x)$
appearing in Eq.\eqref{eq.BKTCorrelationLength} in Section \ref{Sec.Methods} and we take $a$ to be the
square planar Cu lattice constant, thus completely fixing $\xi(x,T)$ for a given $x$ and $T$.
The main features of our results are not crucially dependent either on the specific form of $\xi(x,T)$ used here or on
the particular chosen value of $b'(x)$ (see Appendix \ref{App.CorrelationLength}). The other input 
$\bar{\Delta}(x,T)$ (Appendix \ref{App.FermiArc}) is fixed by enforcing  $T_\mathrm{an}$, obtained
via the relation $\bar{\Delta}(x,T_\mathrm{an})\simeq \mathrm{v}_\mathrm{an}/(\sqrt{2}\xi(x,T_\mathrm{an}))$, to be
close to the ARPES data\cite{JCCampuzano} for the antinodal pseudogap filling temperature $T^*(x)$ i.e. the
pseudogap temperature scale measured in ARPES. Thus $T_\mathrm{an}(x)$ is \emph{preordained} to be
nearly same as $T^*(x)$ in our calculation, Fig.\ref{fig.FermiArc}{\bf c} manifests this fact.

Some other spectral features observed as $T$ decreases through $T_c$ are shown in
Fig.\ref{fig.SpectralFeatures}. Each of the left figure panels (Fig.\ref{fig.SpectralFeatures}{\bf a, c}) exhibits a 
particular spectral property (see below) over the entire FS (see Fig.\ref{fig.EDC}{\bf a}), for a typical $x\simeq 0.11$ at 
several temperatures. In the respective right panels (Fig.\ref{fig.SpectralFeatures}{\bf b, d})
we show the same spectral quantities specifically at the antinode ($\phi=0$) as a function of temperature for various $x$. 
The pseudogap is widely characterized \cite{AKanigel1,AKanigel2} by the function $L(\phi)=1-A(\phi,0)/A(\phi,\Delta)$ that 
quantifies the loss of spectral weight $A(\phi,\omega)$ for energy $\omega=\pm\Delta(\phi)$ at angular position 
$\phi$ (Fig.\ref{fig.EDC}{\bf a}) on the FS. If there is a nonzero gap in the excitation spectrum, and there are no zero 
energy excitations (as in a BCS superconductor at $T=0$) $L(\phi)=1$; if there is no gap (real or pseudo) $L(\phi)=0$. 
Intermediate values (i.e.~$0<L(\phi)<1$) imply a pseudogap. The evolution of $L(\phi)$ as $T$ decreases through $T_c$ 
describes how the pseudogap develops into a real gap. This is shown in Fig.\ref{fig.SpectralFeatures}{\bf a}. It agrees 
qualitatively with that obtained from ARPES \cite{AKanigel2}. Experiments show a nonzero
$A(\mathbf{k},\omega=0)$ (or $L(\phi)<1$) below $T_c$ as well, unlike in our theory where below $T_c$ , $L(\phi)=1$, 
except at $\phi=45^\circ$ where it is zero. 
The difference is presumably due to neglect of some causes of intrinsic spectral line-shape broadening, not included in our 
model, and finite instrumental resolution of ARPES. Above $T_c$ as well, the 
calculated $L(\phi)$ in the pseudogapped portion is less (by about $20\%$) than the experimental values, 
probably due to such effects. In Fig.\ref{fig.SpectralFeatures}{\bf a}, we have (arbitrarily) multiplied our 
results by a related factor ($\sim1.2$) for 
comparison with experiment. The spectral loss at the antinodal point i.e.~$L(\phi=0)\equiv L_\mathrm{an}$ is
exhibited in Fig.\ref{fig.SpectralFeatures}{\bf b} for several values of $x$ as a function of $T$. 
$L_\mathrm{an}$ decreases almost linearly with temperature in line with observations\cite{AKanigel2} and vanishes at 
$T=T_\mathrm{an}$.

                     Another aspect of the spectral function $A(\mathbf{k},\omega)$ is shown in
Fig.\ref{fig.SpectralFeatures}{\bf c}, in which the full width at half maximum (FWHM) of the peak is plotted as a function 
of $\phi$, again for $x\simeq 0.11$. We notice a maximum in it which coincides with the $\phi$ location of the Fermi arc tip. 
Similar features can be deduced from the ARPES spectra\cite{AKanigel2}. We also find in 
Fig.\ref{fig.SpectralFeatures}{\bf d} that while the FWHM of the antinodal 
peak increases substantially with $T$ below $T_c$ in the underdoped case, it does not change much for the
overdoped cuprate, slightly beyond optimal doping. 
This is what is observed\cite{HDing}. Again, the experimental widths 
are seen to be around 20 meV larger than the calculated ones, perhaps due to the same neglect of some
intrinsic and extrinsic quasiparticle lifetime effects mentioned earlier.

                   Possibly the most widely explored property of a superconductor is the energy gap
$\Delta_\mathbf{k}$ or $\Delta(\phi)$. The apparent deviation (`bending') of the gap below $T_c$, 
inferred from ARPES \cite{KTanaka,WSLee}, 
from the $d$-wave form has been the subject of a great deal of current interest, leading to speculation 
that there are two gaps in high-$T_c$ superconductors \cite{AJMillis}. We have obtained the gap  $\Delta_\mathbf{k}$ from the
peak position of the calculated spectral function $A(\mathbf{k},\omega)$ for a fixed $\mathbf{k}$ as mentioned
earlier and conclude from the results 
(see Fig.\ref{fig.PeakStrength}{\bf a}) that the `bending' is due to the coupling of the electron to thermal phase 
fluctuations (`spin waves') below $T_c$. As expected from such an origin, it is large close to $T_c$, and small as
$T\rightarrow 0$. These results confirm that there is only one gap, but that to `uncover' it, effects of
coupling to pair fluctuations (`spin waves') have to be included.
                         
Below $T_c$, because of the LRO in the Cooper pair amplitude, the electrons move in a lattice periodic pair
potential while \emph{occasionally} getting scattered from the thermal `spin wave' fluctuations around the ordered
state, so that the eigenstates have coherent quasiparticle features. 
One consequence, shown in Fig.\ref{fig.PeakStrength}{\bf b}, is that 
the height ($A^p_\mathrm{an}$) of the coherent antinodal peak at $\omega=\pm\Delta_\mathrm{an}$ follows closely the $T$-dependence of 
$\eta^{-1}\propto(\rho_s/T)$. A similar empirical correlation has been reported in Refs.\cite{HDing,DLFeng}. At a given 
temperature, $A^p_\mathrm{an}$ is proportional to $\rho_s(x,T)$. This is shown in
the inset of Fig.\ref{fig.PeakStrength}{\bf b} for a particular value of $T$ as a function of $x$.

\section{Discussion} \label{Sec.Discussion}

Our work is a comprehensive exploration of the consequence of the (inevitable) coupling between electrons and
Cooper pair fluctuations constituted of the same electrons in a two dimensional model for the cuprates with a
square lattice. Underlying this is a picture of the superconducting transition as a continuous approach to
long range order characterized by nonzero phase stiffness so that the pair (phase) correlation length $\xi$
diverges at $T_c$. (One way this can happen, which is the mechanism discussed in an earlier paper
\cite{SBanerjee} by us in a Ginzburg-Landau-like theory of superconductivity in the cuprates, is the
following. Nearest neighbor spin singlet pairs are the basic degree of freedom, and are preformed well above
$T_c$ for optimal and suboptimal hole density. Their nearest neighbor phase dependent interaction of `AF' type
leads to $d$-wave symmetry LRO below $T_c$). 

We have obtained here explicitly the self energy
$\Sigma(\mathbf{k},i\omega_n)$ of an electron moving in the field of Cooper pairs whose $d$-wave symmetry
correlation length $\xi$ diverges as $T\rightarrow T_c^+$. This is done very generally, e.g.~assuming that the
large distance ($R>>a$) phase correlator function is of the form $R^{-\eta}\exp{(-R/\xi)}$, where $\eta$ is
the anomalous dimension. In actual calculations we use the specific forms (and values where appropriate) for
$\eta$ and $\xi$ from the BKT theory \cite{VLBerezinskii,JMKosterlitz1,JMKosterlitz2} without making any
further assumptions regarding $\eta$ and $\xi$. Since the BKT theory predicts somewhat lower $T_c$ compared to
experiment ($T_c-T_\mathrm{BKT}\sim 5-10$ K) and since the effect of small interlayer coupling which leads to
the observed three dimensional (XY) critical behavior very close to $T_c$ is neglected, we compare our results
for spectral function $A(\mathbf{k},\omega)$ with ARPES experiments outside a regime $\delta T\simeq 10-15$ K
from $T_c$. In this paper, we concentrate on ARPES results because ARPES represents real space averaged
measurements. We do not address STM (Scanning Tunnelling Microscopy) results because the latter is a spatially
local measurement sensitive to local inhomogeneities. 

The lowest order vertex correction to the self energy shown in Fig.\ref{fig.BondLattice}{\bf a} vanishes identically above $T_c$. This is because the vertex correction involving the Cooper pair correlation function $\langle \psi_m\psi^*_n\rangle$ requires the anomalous electron propagator to be non zero. The latter  is identically zero since there is no long range superconducting order. The part of the vertex correction involving $\langle \psi^*_m\psi^*_n\rangle$ (or $\langle \psi_m\psi_n\rangle$) is zero because such functions vanish due to `isotropy' or `XY-symmetry' in the space of two-component `spin' $\psi_m=(\Delta_m\cos\theta_m,\Delta_m\sin\theta_m)$. Below $T_c$, the correction can be calculated exactly as in the classic work of Migdal \cite{ABMigdal}. The internal phonon propagator  of the Migdal calculation is replaced here by the collective phase fluctuation or ‘spin’ wave or Goldstone mode propagator, so that the correction is $\sim(T_c/\epsilon_\mathrm{F})\sim 10^{-2}<<1$.  However, higher order vertex corrections are nonzero. These have not been calculated. The internal `bosonic' or pair fluctuation propagator is the true one, not the bare one, since it is described in terms of the observed transition temperature $T_c$ which fully includes dressing or renormalization effects . The approximation made is that the bare fermion (electron/hole) propagator is used in calculating the self energy instead of the dressed one. Experience with the Eliashberg approximation in phonon induced Cooper pair theory (BCS) as well as calculations for coupled electron phonon systems in the normal state show that this is reasonable. We would like to point out that the role of the `boson' here is not to form Cooper pairs (as in the celebrated BCS theory); the spin singlet nearest neighbour Cooper pairs are  already formed while the interaction between them leads to the emergence of d wave symmetry superconductivity (long range order or LRO) below  $T_c$ as a collective effect. The `boson' of Fig.\ref{fig.BondLattice}{\bf a} is the correlation at large distance between $d$-wave symmetry or collective Cooper pair fluctuations, whose length scale diverges at $T_c$.

 We do \emph{not} attempt to explain the broad part of the one
electron Green function \cite{ADamascelli,JCCampuzano,BEdegger}, specially prominent away from the node, both above and 
below $T_c$ (this `incoherent' part may well be a strong correlation effect). We assume however that at low
excitation energies there is a `quasiparticle' part in the `bare' one electron Green function, with weight
$z_\mathbf{k}$ (the corresponding Green function, that we calculate here, only gets scaled by $z_\mathbf{k}$,
i.e.~$G(\mathbf{k},i\omega_n)\rightarrow z_\mathbf{k}G(\mathbf{k},i\omega_n)$, so that the area under the
quasiparticle part of the spectral function $A(\mathbf{k},\omega)$ is $z_\mathbf{k}$ \cite{BEdegger}). In strongly correlated system $z_\mathbf{k}$ is usually $\mathbf{k}$-independent \cite{BEdegger}. Recent QMC simulations \cite{BMoritz,AMacridin} of the one band Hubbard model in two dimensions (widely believed to be appropriate for the cuprates)shows that for strong correlations, there is indeed a quasiparticle part to the one electron Green’s function, and that the quasiparticle residue $z_\mathbf{k}$  is weakly $\mathbf{k}$ dependent. Presumably, such a $z$ can
be absorbed into a definition of renormalized $\bar{\Delta}^2$,
i.e.~$\left(\bar{\Delta}^2\right)_\mathrm{renorm}=(\bar{\Delta}^2z)$ while evaluating $G(\mathbf{k},\omega)$
using the Dyson equation (Eq.\eqref{eq.DysonEquation}) and the self energy expression of
Eq.\eqref{eq.StaticSelfEnergy}, as $G^0(\mathbf{k},\omega)\rightarrow zG^0(\mathbf{k},\omega)$ there. The
electronic self energy we calculate as a result of coupling to pair fluctuations depends strongly on
$\mathbf{k}$; in particular it vanishes at the nodal point.

There have been a large number of calculations of the electron spectral density in cuprates under more specific
assumptions and addressing particular experimental findings, e.g.~the Fermi arc or the pseudogap. We mention
below some that we are aware of. None, as far as we know, compare theoretical results with those of actual
experiments explicitly, over this wide a range of temperature, doping, experiments and phenomena or uses this
picture of the superconducting phase transition.

Many earlier calculations of the effect of phase fluctuations (e.g.~Franz and Millis \cite{MFranz}, Berg and
Altman \cite{EBerg}) assume that these are connected specifically with supercurrents surrounding thermal
vortices. While we use the BKT model, which describes the superconducting transition as due to
vortex-antivortex unbinding, and the associated correlation length $\xi$ (Eq.\eqref{eq.BKTCorrelationLength}),
our form for the self energy (Eq.\eqref{eq.MatsubaraSelfEnergy}) is independent of the actual mechanism for
the phase fluctuation spectrum or the way $\xi$ diverges approaching $T_c$, depending only on the generally
valid form for the spatial dependence of of the correlation function of the phase fluctuations. Some other
recent calculations (Micklitz and Norman \cite{TMicklitz}, Senthil and Lee \cite{TSenthil}) either use
different forms for $D(R)$ or assumes implicitly that $\eta=0$ and $\xi=\infty$. The latter implies preformed
$d$-wave symmetry pairs which is almost the prevailing belief in the field. In our approach, $d$-wave symmetry
long range order develops at $T_c$; above it this begins to \emph{emerge} as a collective phenomenon as
$\xi\rightarrow \infty$. A recent calculation by Tsvelik and Essler \cite{AMTsvelik} is similar in spirit to
ours, but in our opinion it is too strongly tied to the anisotropic Kondo limit. Perhaps the
closest is the recent numerical calculations by Li and coworkers, who have analyzed the electron self energy $\Sigma$ via 
Monte Carlo sampling of 2D-XY model \cite{QHan} as well as the Hubbard-Stratonovich transformed version \cite{YZhong} of the 
pair attraction term of Eq.\eqref{eq.Hamiltonian} using an involved numerical technique. Their results are similar overall 
to some of those described here, e.g. in Fig.\ref{fig.FermiArc}{\bf a}.

Our approach for calculating the low energy excitation spectrum of electrons coupled to quasi static
long-wavelength fluctuations implies that an electron decays into another with \emph{nearly} opposite momentum
and a Cooper pair fluctuations with small momentum $2\mathbf{q}$. This is taken to be the only contribution to
quasiparticle decay near the Fermi energy. We do not obtain or emphasize finite Cooper pair life-time effects,
generally called `dynamic' effects in the literature, and mostly included empirically to fit ARPES data. We
also do not use results from BCS theory which being a mean field theory, can not, we believe, describe
accurately the region above $T_c$ (and say, below $T^*$) since here local Cooper pairs (nearest neighbor spin
singlets) exist, but condense collectively into a state of $d$-wave symmetry LRO only at $T_c$.

As mentioned earlier, a variety of \emph{exotic}
mechanisms, such as stripes \cite{SAKivelson1}, $d$-density wave \cite{SChakravarty} and time reversal
symmetry breaking orbital current state \cite{CMVarma}, have been suggested and invoked to describe the
phenomena observed in the cuprates (especially above $T_c$). Our work provides an explanation in terms of a
simple, inevitable process. We believe that the results, and their agreement with experiments, strongly
support our mechanism \cite{SBanerjee} of nearest neighbor Cooper pairs, and long-range $d$-wave symmetry
order emerging as a collective effect arising from short-range interaction between these pairs. This probably
points to the way in which, we believe, high-$T_c$ superconductivity will be understood.

We do not include here the effect of coupling to additional short-range fluctuations which we find to be
present. These may have a decisive effect on nodal quasiparticle life-time as well as on linear resistivity in
the strange metal phase. We have also not emphasized particle-hole symmetry breaking terms in
$A(\mathbf{k},\omega)$ which we have calculated. Our approach implies that fluctuation effects, very small in
conventional superconductors, are sizable in the cuprates and are present over a wide temperature range.

\begin{acknowledgments}
We thank S.~Mukerjee and U.~Chatterjee for useful discussions. S.B.~would like to acknowledge CSIR
(Govt.~of India) and DST (Govt.~of India) for support. T.V.R.~acknowledges research support from the DST (Govt.~of India) through the Ramanna
Fellowship as well as NCBS, Bangalore for hospitality. C.D.~acknowledges support from DST (Govt.~of India).
\end{acknowledgments}

\appendix

\section{Determination of the Fermi surface} \label{App.FermiSurface}
The chemical potential $\mu$ is computed by setting $\int_{-\infty}^\infty d\omega\sum_{\mathbf{k}} f(\omega)\mathcal{A}(\mathbf{k},\omega)=(1-x)$
(here $f(\omega)=1/(e^{\beta\omega}+1)$ is the Fermi function) and the underlying Fermi surface is defined from the locus of
$\xi_\mathbf{k}=0$. We have implemented other FS criteria in our calculation e.g. the locus of the minimum of 
$\Delta_\mathbf{k}$ \cite{AKanigel1} as well as the locus of the maximum of $A(\mathbf{k},\omega=0)$. The main features of 
the results (Figs.\ref{fig.EDC}-\ref{fig.PeakStrength}) remain unaltered for different FS criteria.

\section{Evaluation of self energy} \label{App.SelfEnergy}
When the pair correlator $D(\mathbf{R})$ in Eq.\eqref{eq.UrsellFunction} is sufficiently long-ranged 
(below $T_c$ at all temperatures, and above $T_c$ for temperatures such that $\xi(T)>>a$), its Fourier transform
$D(2\mathbf{q})=\sum_\mathbf{R} D(\mathbf{R})\exp{(-i2\mathbf{q}.\mathbf{R})}$ is sharply peaked
around $\mathbf{q}=0$ and one can expand $\xi_{\mathbf{k}-2\mathbf{q}}$ and 
$f_\mu(\mathbf{k},\mathbf{q})$ in Eq.\eqref{eq.StaticSelfEnergy} in powers of $\mathbf{q}$, so that 
$\Sigma(\mathbf{k},i\omega_n)$ can be approximated as,
\begin{eqnarray}
\Sigma(\mathbf{k},i\omega_n)&\simeq& \mathcal{R}(\cos{k_xa}-\cos{k_ya})^2+\ldots,
\label{eq.SelfEnergyExpansion}
\label{eq.ExpansionSelfEnergy}
\end{eqnarray}
where
$\mathcal{R}=N^{-1}\sum_\mathbf{q}D(2\mathbf{q})/(i\omega_n+\xi_\mathbf{k}-2\mathbf{v}_\mathbf{k}.\mathbf{q})$.
In Eq.\eqref{eq.SelfEnergyExpansion}, we have not shown explicitly particle-hole non-symmetric contributions to
$\Sigma(\mathbf{k},i\omega_n)$ for which we have obtained closed form expressions. 
This can be recast using standard mathematical identities into forms suitable for analytical and numerical
calculations, i.e.
\begin{eqnarray}
\mathcal{R}&=-\frac{i~\mathrm{sgn}(\omega_n)}{4}\int_0^\infty ds
e^{i\mathrm{sgn}(\omega_n)(i\omega_n+\xi_\mathbf{k})s} D(s\mathrm{v}_\mathbf{k}),\label{eq.SelfEnergyIntegral}
\end{eqnarray}
The above expression utilizes only the 
{\it real-space pair correlator} $D(R)$ (Eq.\eqref{eq.UrsellFunction}). This procedure is very different from the way
self energy is generally calculated using momentum space form of $D(R)$ (see e.g.~Refs.\cite{TSenthil,TMicklitz}). 
For instance, using $D(R)\simeq
\bar{\Delta}^2(T)\bar{D}(R)$ (Eq.\eqref{eq.PairCorrelator}) along with $\bar{D}(R)=(\tilde{\Lambda}R)^{-\eta}\exp{(-R/\xi)}$ 
($\tilde{\Lambda}=e^{-0.116}2\sqrt{2\pi}/a$ is related to upper wave-vector cutoff \cite{PMChaikin}) 
results in the self energy expression of
Eq.\eqref{eq.MatsubaraSelfEnergy}. As evident, the integration variable
$s$ in Eq.\eqref{eq.SelfEnergyIntegral} has dimension of time or inverse energy and hence the presence of 
$D(sv_\mathbf{k})\sim \exp{(-sv_\mathbf{k}/\xi)}$ in Eq.\eqref{eq.MatsubaraSelfEnergy} cuts
off the contributions for large timescales corresponding to $s>>\xi/v_\mathbf{k}$. Below $T_c$, $D(R)$, calculated from a spin 
wave approximation incorporating finite interlayer coupling (see below), is used in Eq.\eqref{eq.SelfEnergyIntegral} to 
obtain $\Sigma(\mathbf{k},i\omega_n)$ by numerical integration. This also demonstrates the usefulness of
Eq.\eqref{eq.SelfEnergyIntegral} for evaluating self energy using arbitrary form of the real-space pair
correlator as it amounts to carrying out a one dimensional integral, either analytically or numerically. 

\begin{figure}
\begin{center}
\begin{tabular}{c}
\includegraphics[height=6cm]{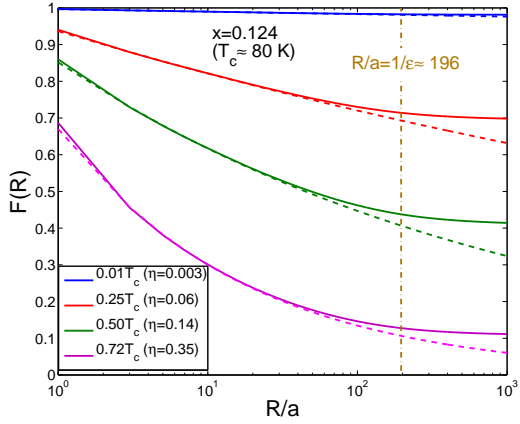}
\end{tabular}
\end{center}
\caption{{\bf Phase correlator $\bar{D}(R)$ below $T_c$.} $\bar{D}(R)$ (solid lines) estimated using
Eq.\eqref{eq.PhaseCorrelator} for $x=0.124$ at four different temperatures. Also shown are the corresponding 2D
correlator $\bar{D}_\mathrm{2D}(R)$ (dashed lines). The vertical dashed line indicates the length scale below which
$\bar{D}(R)\simeq \bar{D}_\mathrm{2D}(R)$.}
\label{fig.PhaseCorrelator}
\end{figure}

\begin{figure*}
\begin{center}
\begin{tabular}{ccc}
\includegraphics[height=6cm]{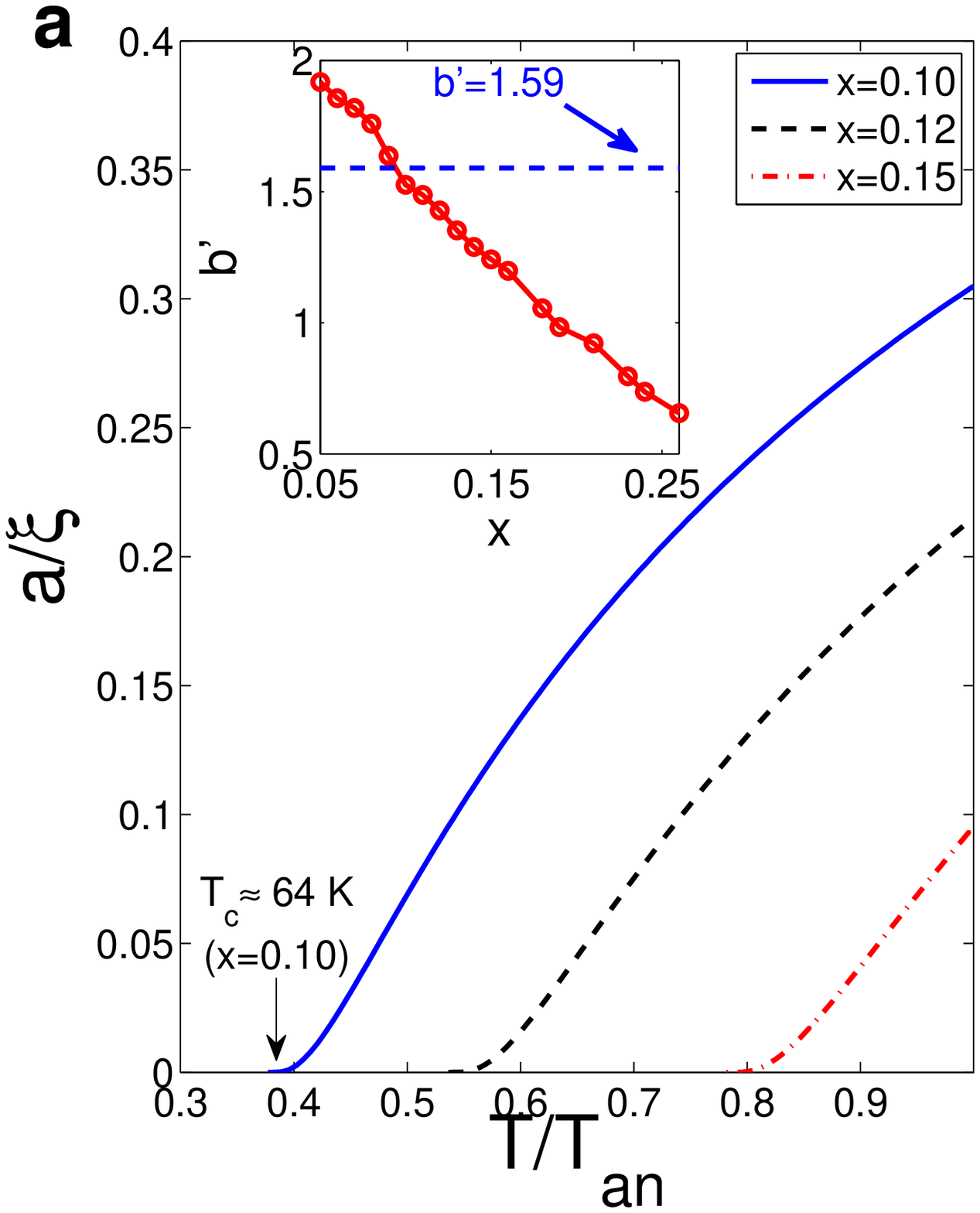}&
\includegraphics[height=6cm]{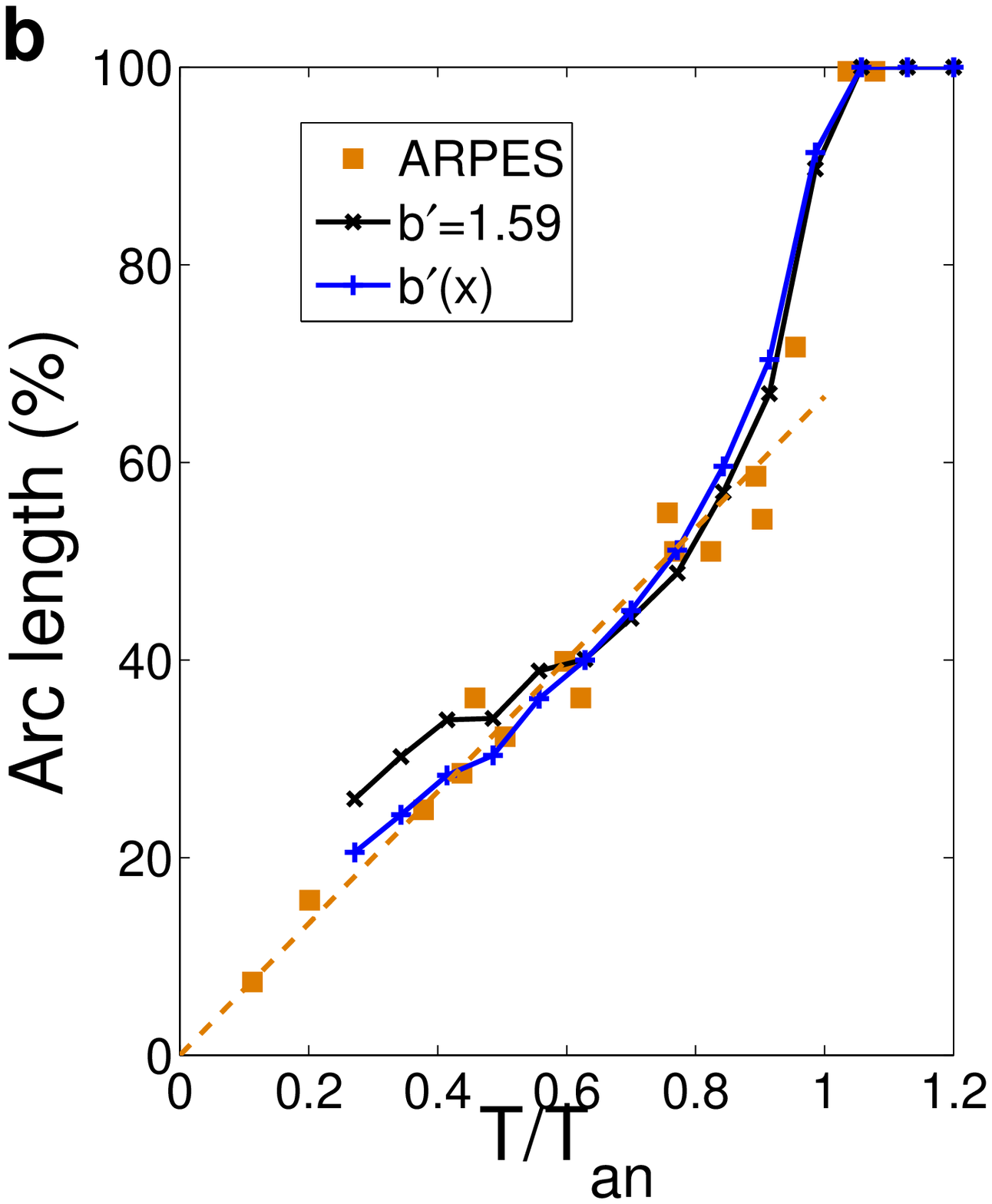}&
\includegraphics[height=6cm]{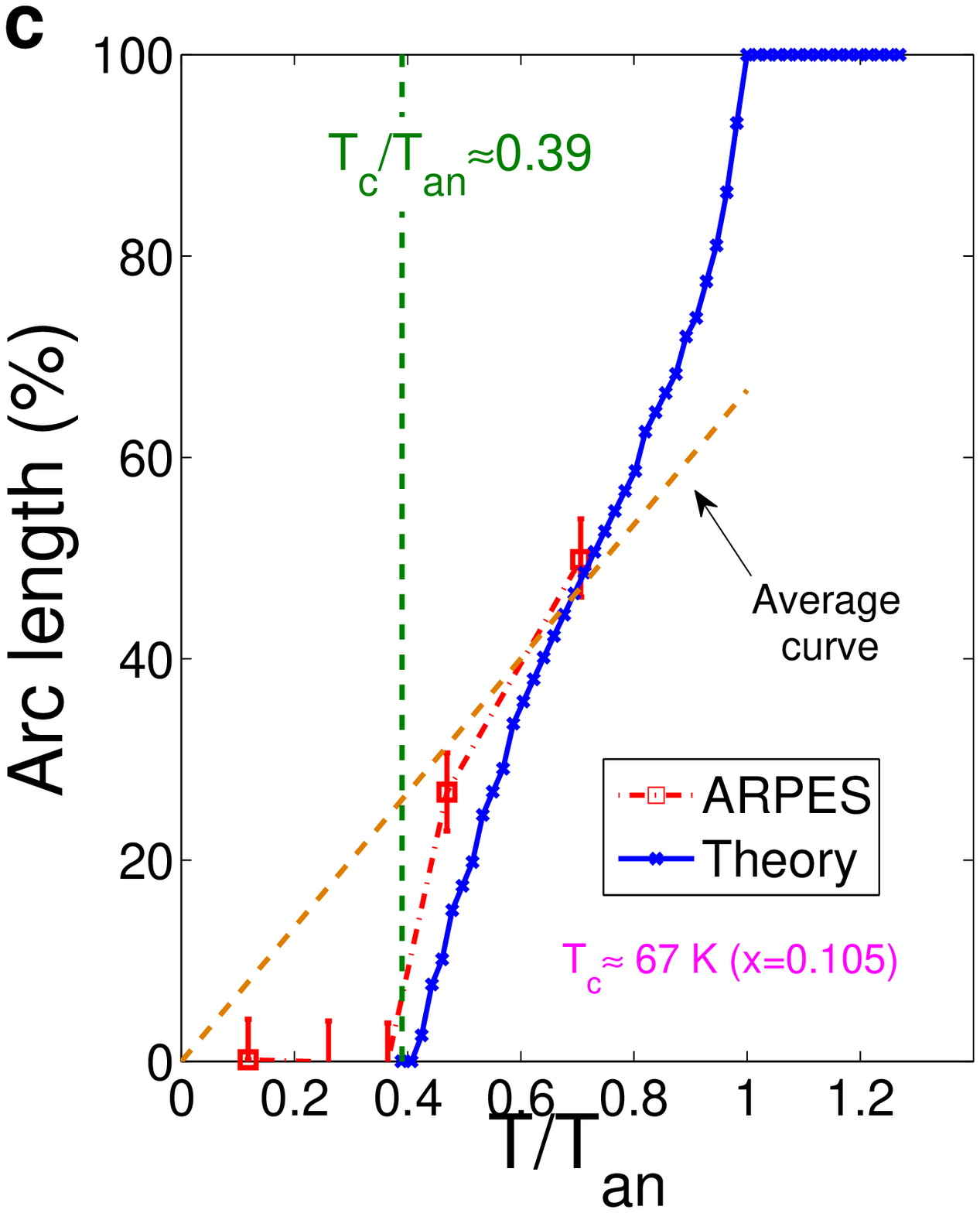}
\end{tabular}
\end{center}
\caption{{\bf Correlation length $\xi(x,T)$ and Fermi arc length.} {\bf a}, The correlation length $\xi$,
which controls the approximate linear temperature dependence of arc length with temperature in
Fig.\ref{fig.FermiArc} {\bf b}, is shown as a function of dimensionless temperature $T/T_\mathrm{an}(x)$ for four 
$x$ values. {\bf Inset:} $b'(x)$ estimated from the GL-like functional of 
Ref.\onlinecite{SBanerjee}. The horizontal dashed line corresponds to $b'=1.59$, as estimated for a 2D-XY model 
\cite{POlsson}. {\bf b}, The arc length vs.~$T/T_\mathrm{an}(x)$ curves (the average curve for
the doping range $0.07\leq x \leq 0.15$) for two different choices of $b'$, namely $b'=1.59$ and an $x$-dependent 
$b'$ estimated using the functional of Ref.\onlinecite{SBanerjee}. The rapid increase of arc length near
$T/T_\mathrm{an}=1$ is related with the rapid decrease of $\langle \Delta_m\rangle=\bar{\Delta}(T)$ near
$T_\mathrm{an}$ (see Fig.\ref{fig.LocalGap}). {\bf c}, The arc length as a function of $T/T_\mathrm{an}$ for $x\simeq 0.11$. ARPES results\cite{AKanigel1} for the arc length for the same doping at a few
values of $T/T_\mathrm{an}$ are also shown.}
\label{fig.CorrelationLength}
\end{figure*}

\section{Pair correlator $D(\mathbf{R})$ below $T_c$ in the presence of
non-zero interlayer coupling} \label{App.PairCorrelator}

\begin{figure}
\begin{center}
\begin{tabular}{cc}
\includegraphics[height=5.7cm]{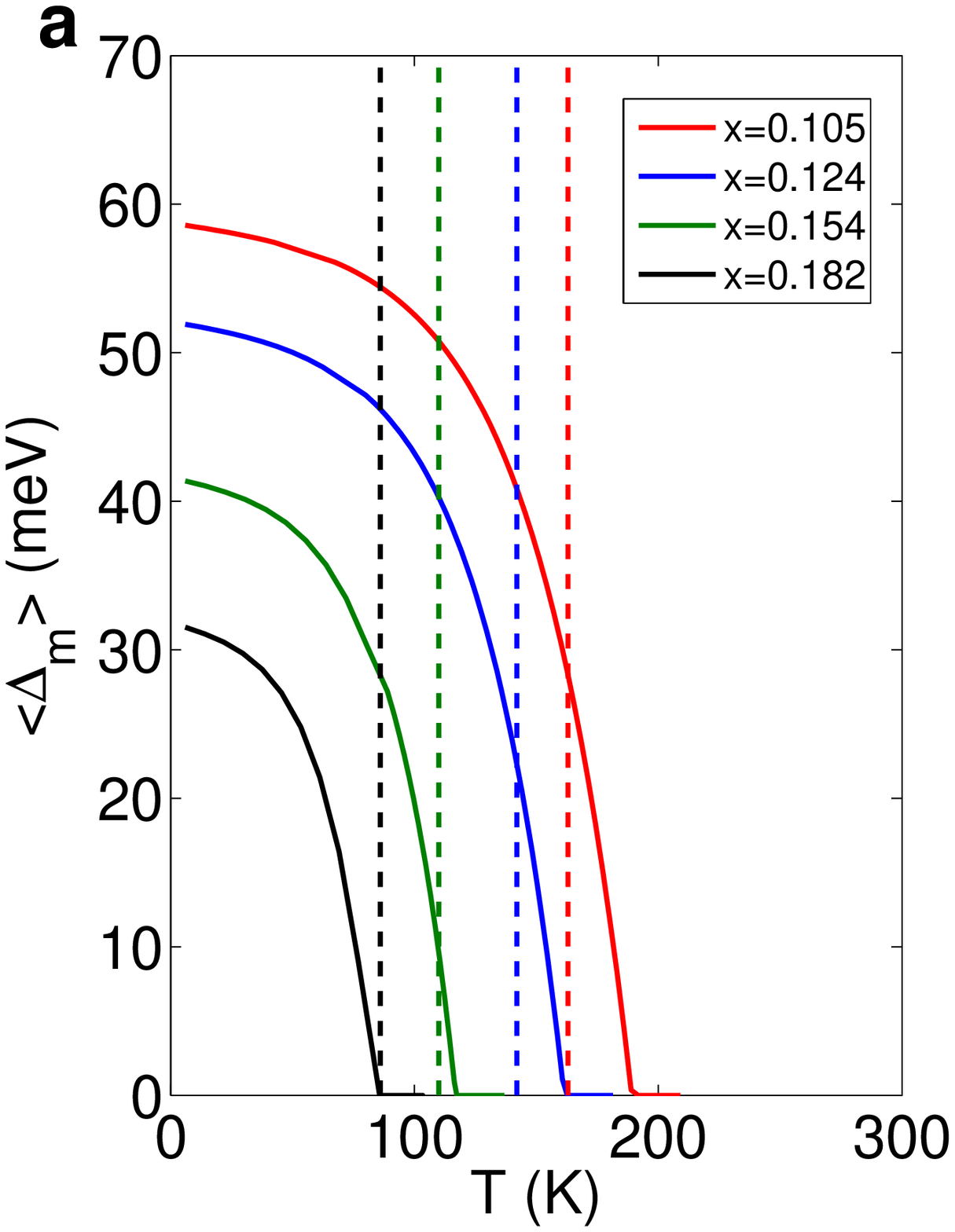}&
\includegraphics[height=5.7cm]{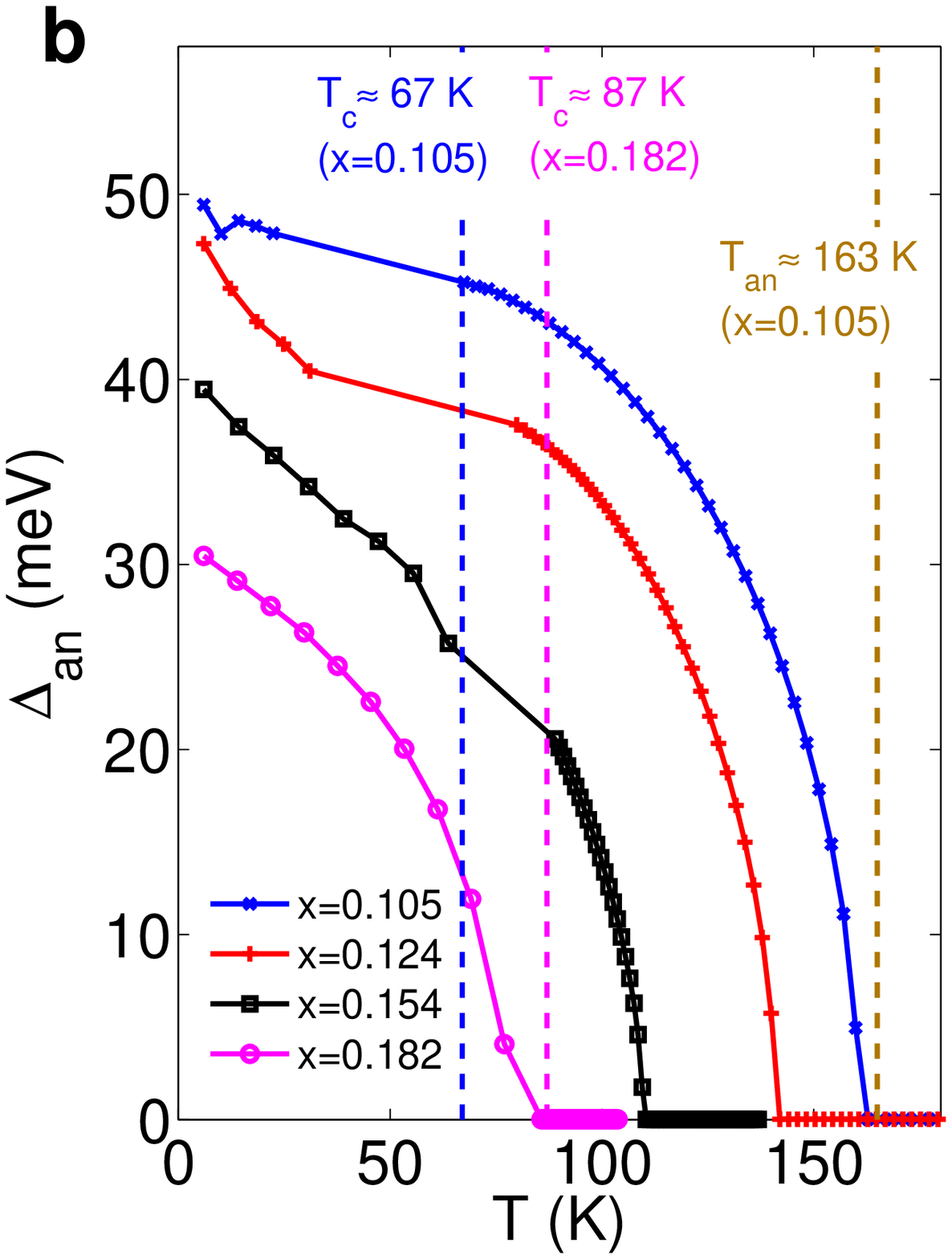}
\end{tabular}
\end{center}
\caption{{\bf Average local gap and antinodal gap.} {\bf a}, $\bar{\Delta}(T)=\langle \Delta_m \rangle$ for four values of hole 
density. Vertical dashed lines indicate the respective antinodal pseudogap filling temperatures $T_\mathrm{an}$. {\bf b}, 
$\Delta_\mathrm{an}(T)$ for the same hole densities.}
\label{fig.LocalGap}
\end{figure}

Below $T_c$, we estimate the phase correlator $\bar{D}(R)=\langle e^{i(\tilde{\theta}(\mathbf{R})-\tilde{\theta}(\mathbf{0}))}\rangle$ of 
Eq.\eqref{eq.UrsellFunction} using the well known harmonic spin wave approximation i.e. the elastic free
energy functional for an anisotropic 3D system,
\begin{eqnarray}
\mathcal{F}_\mathrm{SW}&=&\frac{\rho_s(T)}{2c}\int d\mathbf{r} [(\partial_x\varphi)^2+(\partial_y
\varphi)^2]\nonumber \\
&+& \frac{\rho_s^c(T)c}{2a^2}\int d\mathbf{r} (\partial_z \varphi)^2, \label{eq.SWHamiltonian}
\end{eqnarray}
where $\partial_x\equiv \frac{\partial}{\partial x}$ and $c$ is the interlayer spacing. As mentioned earlier, 
$\rho_s=\rho^{ab}_s$ and $\rho_s^c$ are the $ab$-plane and $c$-axis superfluid density, respectively. The phase correlator 
can be written \cite{PMChaikin} as $\bar{D}(R)=\exp{(-g(R))}$, with
\begin{eqnarray}
&&g(R) \nonumber \\
&&=\frac{Tc}{\rho_s(T)}\int_{-\pi/a}^{\pi/a}\frac{dq_xdq_y}{(2\pi)^2}\int_{-\pi/c}^{\pi/c}\frac{dq_z}{2\pi}
\frac{1-e^{i\mathbf{q}.\mathbf{R}}}{q_x^2+q_y^2+\varepsilon^2(cq_z/a)^2},
\nonumber \\
\end{eqnarray}
 Here $\varepsilon=(\rho_s^c/\rho_s)^{1/2}$ and $\mathbf{R}=(R_x,R_y,0)$ is in the $xy$-plane.
Finally one obtains,
\begin{eqnarray}
&&\bar{D}(R) \nonumber \\
&&=\exp{\left[-\left(g_\mathrm{LRO}-\frac{\eta}{\pi \varepsilon}\int_0^\Lambda dq
J_0\left(\frac{qR}{a}\right)\tan^{-1}\left(\frac{\varepsilon \pi}{q}\right)\right)\right]} \nonumber \\\label{eq.PhaseCorrelator}
\end{eqnarray}
Here, $J_0$ is the ordinary Bessel function, and
$g_\mathrm{LRO}=\frac{\eta}{2\pi\varepsilon}[\Lambda\pi-2\Lambda \cot^{-1}(\varepsilon\pi/\Lambda)+\varepsilon
\pi \ln(1+(\Lambda/\pi \varepsilon)^2)]$ with $\Lambda=2\sqrt{\pi}$ and $\eta=T/(2\pi\rho_s)$. The phase correlator 
$\bar{D}(R)$ is shown in Fig.\ref{fig.PhaseCorrelator}{\bf a} for $x=0.124$ at different temperatures.
$\bar{D}(R)$ starts from $\bar{D}(0)=1$ and decreases to a constant value $\bar{D}_\mathrm{LRO}=\exp{(-g_\mathrm{LRO})}$ as
$R\rightarrow \infty$. The fluctuation part $\widetilde{D}(\mathbf{R})$ (Eq.\eqref{eq.UrsellFunction}) can be obtained 
as $\widetilde{D}(R)\simeq \bar{\Delta}^2(T)(\bar{D}(R)-\bar{D}_\mathrm{LRO})$. In Fig.\ref{fig.PhaseCorrelator}{\bf a}, we also show the 
corresponding 2D phase correlator ($\bar{D}_\mathrm{2D}(R)=(\tilde{\Lambda} R)^{-\eta}$) which is appropriate for a
QLRO phase as $\bar{D}_\mathrm{2D}(R\rightarrow \infty)\rightarrow 0$; $\bar{D}_\mathrm{2D}(R)$ is also obtained from the spin
wave approximation for a 2D system\cite{PMChaikin}. For $a\aplt R<\varepsilon^{-1}a$, $\bar{D}(R)\simeq
\bar{D}_\mathrm{2D}(R)$ i.e. the system effectively behaves as two dimensional. Both the spin wave approximations,
the anisotropic 3D and pure 2D, are expected to be quantitatively correct only at low temperatures.
Especially the estimation of the LRO part of $\bar{D}(R)$ i.e.~$\bar{D}_\mathrm{LRO}$ obtained using the anisotropic 3D
calculation is only accurate at very low temperature for large anisotropy or small $\varepsilon$. In our
calculation, this fact is manifestly observed e.g.~in the variation of the antinodal gap $\Delta_\mathrm{an}$ with
temperature for $T<T_c$ (see Fig.\ref{fig.LocalGap}{\bf b}). The planar 
superfluid density $\rho_s(T)$ is estimated from the measured $ab$-plane penetration depth \cite{WAnukool} for Bi2212 using
$\rho_s=\Phi_0^2c/(16\pi^3\lambda^2_{ab})$; $\Phi_0$ is the fundamental flux quantum and we take
$c=15~\AA$ as suitable for Bi2212. For the anisotropy ratio $\varepsilon$ (assumed to be temperature independent), we use the
empirical formula \cite{TSchneider} $\varepsilon(x)=\varepsilon(x_\mathrm{opt})(x-x_u)/(x_\mathrm{opt}-x_u)$, where 
$x_\mathrm{opt}=0.16$ is the optimal hole doping and $x_u=0.047$ is the end point of the $T_c(x)$ dome in the underdoped side,
with $\varepsilon(x_\mathrm{opt})\simeq 1/133$.

\section{Correlation length $\xi(x,T)$ above $T_c$} \label{App.CorrelationLength}

As mentioned in Section \ref{Sec.Methods}, $\xi(x,T)$ is estimated using the BKT form \cite{PMChaikin} i.e.~$\xi(x,T)\simeq a
\exp{(b'(x)/\sqrt{T/T_c(x)-1})}$. To obtain the actual correlation length, one needs the value of the dimensionless 
quantity $b'(x)$, which is in the exponent. This is a nonuniversal number of order 1. For thin superconducting films its 
value ranges \cite{GBlatter} from 1 to 4. Unfortunately for cuprates no such estimate exists for $b'$ (presumably doping and 
material dependent). We have estimated $b'$ using a GL-like functional in
Ref.\onlinecite{SBanerjee}. As shown there, with appropriate choices of a few parameters as input in the
GL-like functional, it can 
be used to explain and reconcile a wide variety of experimental data for the cuprates e.g.~those for $T_c(x)$,
$\rho_s(x,T)$, and the contribution of pair degrees of freedom to the electronic specific heat.
The quantity $b'(x)$ has been estimated there in the following manner. In two 
dimensions, the system described by the model undergoes a BKT transition at a temperature $T_\mathrm{BKT}(x)$. So, 
the correlation length $\xi(x,T)$ is expected to follow the aforementioned BKT form above $T_c(x)$, where
$T_c(x)$ is identified with $T_\mathrm{BKT}(x)$. In such a scenario, as predicted by BKT theory
\cite{PMChaikin}, the superfluid density  below $T_c(x)$ is given by the formula
\cite{PMChaikin,VAmbegaokar} $\rho_s(x,T)=\rho_s(T_c(x))[1+b(x)\sqrt{1-T/T_c(x)}]$ near the transition.
The BKT theory further predicts a universal relation\cite{VAmbegaokar} between $b'$ and $b$, namely
$bb'=\pi/2$. Keeping this fact in mind, we calculate $\rho_s(x,T)$ below $T_c$ from a Monte Carlo simulation of
the GL model\cite{SBanerjee} and fit the results to the above mentioned BKT form for $\rho_s(x,T)$ to
estimate $b(x)$ and thence $b'(x)$. The results for  $b'(x)$ are shown in the inset of Fig.\ref{fig.CorrelationLength}{\bf a}). 

We show $\xi(T)$, estimated 
using $b'(x)$ obtained in this manner, in Fig.\ref{fig.CorrelationLength}{\bf a} for 
four different $x$ values. These results for 
$\xi(x,T)$ have been used in Eq.\eqref{eq.MatsubaraSelfEnergy} to calculate the spectral properties
above $T_c(x)$, as reported in Figs.\ref{fig.EDC}-\ref{fig.PeakStrength} of the main paper. We have also used the constant 
value $b'(x)=1.59$, expected for a 2D-XY model\cite{POlsson}, to repeat the same calculations. The results in the two 
cases only differ in details and their main features over the entire $x$ range are found to be robust with respect to 
different choices of $b'(x)\sim 1$ (see Fig.\ref{fig.CorrelationLength}{\bf b}). These temperature 
dependence of $\xi^{-1}$ (along with $\langle \Delta_m \rangle =\bar{\Delta}(T)$, 
see below) governs the behaviour of the arc length as a function of $T$ (see Fig.\ref{fig.CorrelationLength}{\bf c}). Starting from zero at $T_c$, the arc length rises steeply with $T$ in a temperature range ($\sim 15$ K) from $T_c$ (the actual value
depending on $x$) to ultimately follow the mean curve shown in Fig.\ref{fig.CorrelationLength}{\bf b}. In our calculation 
this temperature range depends primarily on the details of the temperature
dependence of $\xi$ close to $T_c$ and hence the range can be tuned by changing $b'$ as well as the $T_c$
appearing in Eq.\eqref{eq.BKTCorrelationLength}. 

We use actual observed $T_c$ values for evaluating $\xi$ from Eq.\eqref{eq.BKTCorrelationLength}. It is well
known that the BKT theory result for $T_c$ ($T_\mathrm{BKT}$) in principle can be substantially lower than the
observed $T_c$ \cite{GBlatter}. Also, we have neglected interlayer coupling which affects $\xi(T)$ and indeed
determines the critical behavior very close to $T_c$ (It is 3D-XY \cite{TSchneider} and this 3D critical
regime is about a degree or two wide). Because of the two reasons, we compare our results for the Fermi arc
length in Fig.\ref{fig.FermiArc} with experiment only beyond a regime $\delta T\sim 10-15$ K from $T_c$.
 Using our theory, $\xi(x,T)$ itself can be estimated more accurately at a phenomenological level from a detailed 
comparison with experimental data for the arc length as a function of $T$, especially in a temperature region close to 
$T_c$ since $\bar{\Delta}(T)$ is weakly temperature dependent in this regime.

\section{Fermi arc criterion, average local gap $\bar{\Delta}(x,T)$ and antinodal gap
$\Delta_\mathrm{an}(x,T)$} \label{App.FermiArc}

As already discussed in Section \ref{Sec.Methods}, the genesis of the Fermi arc above $T_c$ can be understood from the 
self energy of Eq.\eqref{eq.MatsubaraSelfEnergy}. Neglecting the small $\eta$ in Eq.\eqref{eq.MatsubaraSelfEnergy},
$\Sigma(\mathbf{k},\omega)=\Sigma(\mathbf{k},i\omega_n\rightarrow \omega+i\delta)$ can be written in a simple
form depicted in Eq.\eqref{eq.SelfEnergy_eta0}. This is exactly same as the phenomenological form\cite{MRNorman2,MRNorman3} 
that has been used widely to analyze the ARPES data using a single particle scattering rate $\Gamma_1$ and $d$-wave pairs 
with finite life-time $\Gamma_0$. In Eq.\eqref{eq.SelfEnergy_eta0}, $\Gamma_0$ is replaced by 
$\mathrm{v}_\mathbf{k}/\xi$ (the single particle scattering rate is zero for $\eta=0$), although our picture
for its origin is completely different, as described in the main paper. 

For $\mathbf{k}$ lying on the Fermi
surface ($\xi_\mathbf{k}=0$), the spectral density $A(\mathbf{k},\omega)$ has a peak at $\omega=0$ for $\mathbf{k}$ values 
such that $|\bar{\Delta}_\mathbf{k}(T)|\leq \mathrm{v}_\mathbf{k}/(\sqrt{2}\xi(T))$ and two peaks at $\omega=\pm
\sqrt{\bar{\Delta}_\mathbf{k}^2(T)-(\mathrm{v}_\mathbf{k}/(\sqrt{2}\xi(T)))^2}$ otherwise. The former determines the 
extended region corresponding to the Fermi arc. The antinodal pseudogap $\Delta_\mathrm{an}\simeq
\sqrt{\bar{\Delta}_\mathrm{an}^2-(\mathrm{v}_\mathrm{an}/(\sqrt{2}\xi))^2}$ is
completely filled in above a temperature $T_\mathrm{an}$, at which the end of the arc reaches the antinodal Fermi
momentum \cite{Footnote1} $\mathbf{k}_\mathrm{an}$ i.e. the criterion of Eq.\eqref{eq.PseudogapFilling} is
satisfied.

At this point, we would like to reemphasize that both $\bar{\Delta}(x,T)$ and $\xi(x,T)$ are purely phenomenological
inputs to our theory. As mentioned in Appendix \ref{App.CorrelationLength}, $\xi(x,T)$ has been determined following fairly
general considerations related to the superconducting transition at $T_c(x)$ (we use the empirical $T_c(x)$
curve \cite{MRPersland} appropriate for Bi2212, see Fig.\ref{fig.FermiArc}{\bf c}). To choose
$\bar{\Delta}(x,T)$, we first \emph{demand}
$T_\mathrm{an}(x)$ to be close to the pseudogap filling temperature $T^*$ obtained in ARPES experiments
\cite{AKanigel2,JCCampuzano}. Since $\mathrm{v}_\mathrm{an}(x)$ (calculated from the
tight-binding energy dispersion $\xi_\mathbf{k}$, see Section \ref{Sec.Methods}) and $\xi(x,T_\mathrm{an})$
(Appendix \ref{App.CorrelationLength}) are already fixed, Eq.\eqref{eq.PseudogapFilling} enables us to deduce
the value of $\bar{\Delta}(x,T_\mathrm{an})$. Also, at $T\simeq 0$, $\bar{\Delta}(x,0)\simeq
\Delta_\mathrm{an}(x,0)$. Using this relation, we obtain the zero temperature antinodal gap
$\Delta_\mathrm{an}(x,0)$, again from the ARPES data\cite{JCCampuzano}, to estimate the value of $\bar{\Delta}(x,T)$ at 
$T=0$. Once both $\bar{\Delta}(x,0)$ and $\bar{\Delta}(x,T_\mathrm{an})$ are known, a simple interpolation
formula, $\bar{\Delta}(x,T)=\bar{\Delta}(x,0)[1-\sinh{(\alpha(x)T)}]$, is used to obtain the parameter
$\alpha(x)$ so that the function $\bar{\Delta}(x,T)$ is completely determined (Fig.\ref{fig.LocalGap}). 
This interpolating form is 
chosen to qualitatively mimic the fact that $\Delta_\mathrm{an}(T)~(\sim \bar{\Delta}(T))$ varies
substantially with temperature only near $T=T_\mathrm{an}$, particularly in the underdoped Bi2212, as reported
in Ref.\onlinecite{AKanigel2}. However, we note that very different temperature variation of $\Delta_\mathrm{an}(T)$, i.e.
$\Delta_\mathrm{an}(T)$ varying considerably with temperature below $T_\mathrm{an}$, has been reported in ARPES
studies \cite{MHashimoto,TYoshida} on other cuprates (e.g.~Bi2201, La214). These observations suggest that the 
form of $\bar{\Delta}(x,T)$
may differ substantially from one material to other. Since $\bar{\Delta}(x,T)$ is a phenomenological input to our theory,
such issues can not be resolved unless a satisfactory microscopic theory for cuprate superconductivity is
developed such that $\bar{\Delta}(x,T)$ is calculable starting from an appropriate microscopic Hamiltonian.

\end{document}